\documentclass[twocolumn,prb,showpacs]{revtex4}
\usepackage{graphicx}
\usepackage{rotating}
\begin{document}
\preprint{ Not ready for distribution }
\title{ Theory for superfluidity in a Bose system }
\author{Zhidong Hao}
\affiliation{ Department of Physics, University of Science and Technology of China, Hefei, Anhui 230026, China }
\date{ \today }
\begin{abstract}
We present a microscopic theory for superfluidity in an interacting many-particle Bose system (such as liquid
$^4$He).  We show that, similar to superconductivity in superconductors, superfluidity in a Bose system arises from
pairing of particles of opposite momenta.  We show the existence of an energy gap in single-particle excitation
spectrum in the superfluid state and the existence of a specific heat jump at the superfluid transition. We derive an
expression for superfluid particle density $n_s$ as a function of temperature $T$ and superfluid velocity ${\bf
v}_s$. We show that superfluid-state free energy density $F$ is an increasing function of $v_s$ (i.e., $\partial
F/\partial v_s > 0$), which indicates that a superfluid has a tendency to remain motionless (this result
qualitatively explains the Hess-Fairbank effect, which is analogous to the Meissner effect in superconductors). We
further speculate the existence of the equation ${\bf j}=-\Lambda\nabla\times \text{\boldmath $\omega$}$, where ${\bf
j} = n_s{\bf v}_s$ is the superfluid current density, $\text{\boldmath $\omega$}=\nabla\times {\bf v}_s$ the
superfluid vorticity, and $\Lambda$ a positive constant (with the help of this equation, the Hess-Fairbank effect can
be quantitatively described).
\end{abstract}
\pacs{67.25.D-} \maketitle

\section{introduction}

We present in this paper a microscopic theory for superfluidity in an interacting many-particle Bose system such as
liquid $^4$He.  The theory is based on an assumption that particles of opposite momenta are paired in the superfluid
state, and thus, is similar in many respects to the BCS theory of superconductivity.\cite{bcs}

It is well known that there is a marked similarity between liquid $^4$He II (the superfluid phase of liquid $^4$He)
and superconductors, both being chiefly characterized by their ability to sustain flows of particles at a constant
velocity without a driving force.\cite{londonI,londonII}  However, unlike superconductors, for which there exists a
successful microscopic theory, i.e., the BCS theory of superconductivity,\cite{bcs} a satisfactory microscopic theory
for liquid $^4$He II is still lacking, despite many efforts (for example, Refs.
\onlinecite{london1938,tisza,landau,bogo1947,feynman}).

Fundamental to the BCS theory of superconductivity is an assumption that electrons of opposite momenta and spins are
paired in the superconducting state.\cite{bcs}  This assumption allows microscopic derivation of all essential
properties of the superconducting state, such as the existence of an energy gap in electronic excitation spectrum, a
second-order phase transition (manifested by a specific heat jump at the superconducting transition), the Meissner
effect, and the Josephson effect.

In this paper we show that it is also the pairing of particles of opposite momenta that is responsible for
superfluidity in a Bose system. Namely, the cause for superconductivity in superconductors and superfluidity in
liquid $^4$He II is indeed essentially the same, irrespective of the nature of the particles involved.

Some previous attempts to develop a microscopic theory for superfluidity in liquid $^4$He II failed at the very start
by assuming that the ground state of liquid $^4$He II is a Bose-Einstein condensate (for example, Ref.
\onlinecite{bogo1947}).  As we will see in this paper, the ground state of a superfluid is not a Bose-Einstein
condensate, but a state in which particles of opposite momenta are paired, similar to that of superconductors.

Pairing of particles in a Bose system has been studied by a number of authors (for example, Refs.
\onlinecite{vb,ei,ns,yinlan}).  However, the authors did not treat properly self-consistency associated with pairing
approximation, and thus, failed to establish a connection between pairing and superfluidity.

In Sec. \ref{secLambda}, we present the theory for the case where a superfluid is at rest, and show the existence of
an energy gap in single-particle excitation spectrum in the superfluid state, and the existence of a specific heat
jump at the superfluid transition. In Sec. \ref{secSuperfluidity}, we present the theory for the case where a
superfluid current is present.  We derive an expression for the superfluid particle density as a function of
temperature and superfluid velocity. We show that the superfluid-state free energy density is an increasing function
of superfluid velocity, which indicates that a superfluid has a tendency to remain motionless. This result provides a
qualitative explanation for the Hess-Fairbank experiment\cite{hess-fairbank} in which a reduction of moment of
inertia was observed when a rotating cylinder of liquid $^4$He was cooled through the superfluid transition (this
phenomenon, known in the literature as the Hess-Fairbank effect, is analogous to the Meissner effect in
superconductors). We further consider how the Hess-Fairbank effect can be quantitatively described. A brief summary
is given in Sec. \ref{secSummary}.

\section{superfluid transition }
\label{secLambda}

We consider an interacting many-particle Bose system.  We assume in this section that superfluid velocity ${\bf
v}_s=0$ (we will consider the case where ${\bf v}_s\neq 0$ in the next section).

Similar to the pairing Hamiltonian in the BCS theory of superconductivity,\cite{bcs} we write the Hamiltonian of the
interacting many-particle Bose system as
\begin{equation}
\hat{H}=\sum_{\bf k}\left(\epsilon_{\bf k}-\mu\right)a^{\dagger}_{\bf k}a_{\bf k}+\frac{1}{2}\sum_{\bf kk'}V_{\bf
kk'}a^{\dagger}_{\bf k}a^{\dagger}_{\bf -k}a_{-\bf k'}a_{\bf k'} , \label{H0}
\end{equation}
where $\epsilon_{\bf k}$ is the normal-state single-particle energy, $\mu$ the chemical potential, $V_{\bf kk'}$ the
pairing interaction matrix element, and $a^\dagger_{\bf k}$ and $a_{\bf k}$ are Bose operators for a single-particle
state of wave-vector {\bf k} in the normal state and satisfy the commutation rule $[a_{\bf k},a^{\dagger}_{\bf
k'}]=\delta_{\bf k,k'}$.

This Hamiltonian can be diagonalized in essentially the same manner as in the BCS theory.\cite{bcs,bogo,valatin}
Namely, we assume
\begin{equation}
\langle a_{-\bf k}a_{\bf k} \rangle \neq 0  \label{pairing}
\end{equation}
in the superfluid state for a pair of ({\bf k}) and $(-{\bf k})$ particles (where the angle brackets
$\langle\cdots\rangle$ denote a thermal average); treat $(a_{-\bf k}a_{\bf k}-\langle a_{-\bf k}a_{\bf k} \rangle)$
as a small quantity so that terms bilinear in $(a_{-\bf k}a_{\bf k}-\langle a_{-\bf k}a_{\bf k} \rangle)$ can be
neglected; define an energy gap parameter
\begin{equation}
\Delta_{\bf k}=-\sum_{\bf k'}V_{\bf k,k'}\langle a_{-\bf k'}a_{\bf k'}\rangle \label{gap-def}
\end{equation}
(because of the similarity between the present theory and the BCS theory of superconductivity, we will similarly
refer to the quantity $\Delta_{\bf k}$ as an ``energy gap parameter'' in this paper, although, as we will see below,
it does not directly relate to an ``energy gap'' in the present theory); and apply a canonical
transformation\cite{bogo1947,bogo,valatin}
\begin{equation}
\label{bogoXform}
\left(
\begin{array}{c}
a_{\bf k} \\ a^{\dagger}_{-\bf k}
\end{array}
\right) = \left(
\begin{array}{cc}
u_{\bf k}   &  v_{\bf k} \\
v^{\star}_{\bf k}  &  u^{\star}_{\bf k}
\end{array}
\right) \left(
\begin{array}{c}
\alpha_{\bf k} \\  \alpha^{\dagger}_{-\bf k}
\end{array}
\right),
\end{equation}
where $\alpha^\dagger_{\bf k}$ and $\alpha_{\bf k}$ are new Bose operators for a single-particle excitation of
wave-vector {\bf k} in the superfluid state, and coefficients $u_{\bf k}$ and $v_{\bf k}$ are so determined as to
diagonalize the Hamiltonian while maintaining the commutation rule $[\alpha_{\bf k},\alpha^\dagger_{\bf
k'}]=\delta_{\bf k,k'}$.

The diagonalized Hamiltonian is
\begin{equation}
{\hat H}=\sum_{\bf k}\left(U_{\bf k}+E_{\bf k}\alpha^{\dagger}_{\bf k}\alpha_{\bf k}\right), \label{H}
\end{equation}
where
\begin{equation}
U_{\bf k} = -\frac{\xi_{\bf k}-E_{\bf k}}{2}+\frac{|\Delta_{\bf k}|^2}{4E_{\bf k}}\left(1+2n_{\bf k}\right);
\label{Uk}
\end{equation}
\begin{equation}
\xi_{\bf k} = \epsilon_{\bf k} - \mu \label{xik}
\end{equation}
is the single-particle energy in the normal state, measured relative to chemical potential $\mu$;
\begin{equation}
E_{\bf k}=\sqrt{\xi_{\bf k}^2-|\Delta_{\bf k}|^2 } \label{Ek}
\end{equation}
the single-particle excitation energy in the superfluid state; and
\begin{equation}
n_{\bf k} =  \left(e^{E_{\bf k}/k_BT}-1 \right)^{-1}
\end{equation}
the Bose function (the number of single-particle excitations of wave-vector {\bf k}).

Coefficients $u_{\bf k}$ and $v_{\bf k}$ are found to satisfy the following relations:
\begin{equation}
|u_{\bf k}|^2 = \frac{1}{2}\left(\frac{\xi_{\bf k}}{E_{\bf k}}+1\right),
\end{equation}
\begin{equation}
|v_{\bf k}|^2 = \frac{1}{2}\left(\frac{\xi_{\bf k}}{E_{\bf k}}-1\right),
\end{equation}
and
\begin{equation}
\Delta_{\bf k}u_{\bf k}v^{\star}_{\bf k} = \frac{ |\Delta_{\bf k}|^2 } { 2 E_{\bf k} }.
\end{equation}

After the diagonalization of Hamiltonian $\hat{H}$, Eq. (\ref{gap-def}) can be expressed as
\begin{equation}
\Delta_{\bf k} = \mbox{} - \sum_{\bf k'} V_{\bf k,k'} \frac{1+2n_{\bf k'}} { 2 E_{\bf k'} } \Delta_{\bf k'}.
\label{gap-eq}
\end{equation}
This is a self-consistency equation that must be satisfied by $\Delta_{\bf k}$ as a function of wave-vector {\bf k}
and temperature $T$.

\subsection{ Critical temperature $T_c$}
\label{subsecTc}

Similar to that in the BCS theory of superconductivity, energy gap parameter $\Delta_{\bf k}$ is an important
quantity in the present theory. It is because of the existence of $\Delta_{\bf k}$ that makes the superfluid state
different from the normal state. In this and the next subsections we consider determination of $\Delta_{\bf k}$.

First, in the limit of $T\rightarrow T_c$ (because of the similarity between the present theory and the BCS theory of
superconductivity, we are similarly using $T_c$, instead of $T_\lambda$, to denote the critical temperature of the
superfluid transition), we have $|\Delta_{\bf k}|\rightarrow 0$ so that Eq. (\ref{gap-eq}) can be linearized and we
have an eigenvalue problem:
\begin{equation}
\Delta_{\bf k} = - \sum_{{\bf k}'} V_{{\bf k},{\bf k}'} \frac{\coth\left[(\epsilon_{\bf
k}-\mu_0)/2k_BT_c\right]}{2(\epsilon_{\bf k}-\mu_0)} \Delta_{{\bf k}'} \,, \label{tc-eq}
\end{equation}
where $\mu_0$ is the value of chemical potential $\mu$ at critical temperature $T_c$, and we have used $1+2n_{\bf
k}=\coth(E_{\bf k}/2k_BT)$ and $E_{\bf k}=\epsilon_{\bf k}-\mu_0$ at $T=T_c$.

Critical temperature $T_c$ and phase $\theta_{\bf k}$ (as defined via $\Delta_{\bf k}=|\Delta_{\bf k}|e^{i\theta_{\bf
k}}$) are determined by solving Eq. (\ref{tc-eq}) for given interaction $V_{\bf k,k'}$, single-particle energy
spectrum $\epsilon_{\bf k}$ and chemical potential $\mu_0$.

Note that it is not necessary to assume $V_{{\bf k},{\bf k'}}<0$ in order for Eq. (\ref{tc-eq}) to have a $T_c>0$
solution. Therefore, the view that an attractive interaction is responsible for particle pairing is incorrect.  Here
we also emphasize that the pairing of particles of opposite momenta, as expressed by Eq. (\ref{pairing}), is a kind
of ordering in momentum space (this point agrees with London's view that superconducting/superfluid state is an
ordered state in momentum space\cite{londonI,londonII}); it does not mean that bound pairs of particles (due to an
attractive interaction) are formed.

\subsection{ $|\Delta_{\bf k}|$ and $E_{\bf k}$ }
\label{subsecHaoEq}

With respect to determination of $|\Delta_{\bf k}|$, we note that the self-consistency equation, Eq. (\ref{gap-eq}),
can be converted into
\begin{equation}
\sum_{\bf k} |\Delta_{\bf k}|^2\frac{\partial}{\partial T}\left(\frac{1+2n_{\bf k}}{E_{\bf k}}\right) = 0
\label{gap-eq11}
\end{equation}
by first operating $\partial /\partial T$ on Eq. (\ref{gap-eq}), and then, multiplying the resulting equation by
$\Delta^{\star}_{\bf k}(1+2n_{\bf k})/2E_{\bf k}$ and summing over {\bf k}.

Interaction $V_{\bf kk'}$ no longer appears in Eq. (\ref{gap-eq11}), because all information about $V_{\bf k,k'}$ is
already contained in $T_c$, and the latter is involved through the condition $|\Delta_{\bf k}(T_c)|=0$.

From Eq. (\ref{gap-eq11}) we can see that the self-consistency equation alone does not allow unique determination of
$|\Delta_{\bf k}|$, because, as one can see,  Eq. (\ref{gap-eq11}) can have an infinite number of solutions. This
property of the self-consistency equation is true for arbitrary interaction $V_{\bf kk'}$, because Eq.
(\ref{gap-eq11}) is derived for arbitrary $V_{\bf kk'}$.  Actually, this property can also been seen directly from
Eq. (\ref{gap-eq}) by noticing that the equation is linear with respect to $e^{i\theta_{\bf k}}$. [Even in the case
of $V_{\bf kk'}=-V$, which leads to $\Delta_{\bf k}=V\sum_{\bf k'}[(1+2n_{\bf k'})/2E_{\bf k'}]\Delta_{\bf k'}$,
$|\Delta_{\bf k}|$ still cannot be uniquely determined, because phase $\theta_{\bf k'}$ of $\Delta_{\bf k'}$ in the
summation over ${\bf k'}$ is measured relative to phase $\theta_{\bf k}$ of $\Delta_{\bf k}$ on the left-hand side of
the equation (it is relative phases that matter). This is more clearly seen if we re-write the equation as
$|\Delta_{\bf k}|=V\sum_{\bf k'}e^{i(\theta_{\bf k'}-\theta_{\bf k})}[(1+2n_{\bf k'})/2E_{\bf k'}]|\Delta_{\bf
k'}|=V\sum_{\bf k'}\cos(\theta_{\bf k'}-\theta_{\bf k})[(1+2n_{\bf k'})/2E_{\bf k'}]|\Delta_{\bf k'}|$, where
$\cos(\theta_{\bf k'}-\theta_{\bf k})=1$ or $-1$ depending on $\theta_{\bf k'}-\theta_{\bf k}=0$ or $\pi$. We have
$|\Delta_{\bf k}|=\Delta={\bf k}\text{-independent}$ only when $\theta_{\bf k}=\theta={\bf k}\text{-independent}$,
but other solutions for $|\Delta_{\bf k}|$, with $\theta_{\bf k}$ being {\bf k}-dependent, are also possible.
Similarly, for a separable interaction of the form $V_{\bf kk'}=-V\omega_{\bf k}\omega_{\bf k'}$, the solution
$|\Delta_{\bf k}|=\Delta|\omega_{\bf k}|$ corresponds to a solution with $e^{i\theta_{\bf k}}=\text{sgn}(\omega_{\bf
k})e^{i\theta}$ with $\theta$ being an arbitrary constant, and is only one of an infinite number of possible
solutions.]

On the other hand, we note that diagonalized Hamiltonian ${\hat H}$ is $T$-dependent, i.e., both $U_{\bf k}$ and
$E_{\bf k}$ in Eq. (\ref{H}) are functions of temperature $T$, because of their dependence upon $|\Delta_{\bf
k}(T)|$.  Since diagonalized Hamiltonian  ${\hat H}$ describes a set of independent excitations, and there is no
transition between different single-particle states in thermodynamic equilibrium, we expect the thermal energy and
entropy associated with a single-particle state of wave-vector {\bf k} to be
\begin{equation}
\varepsilon_{\bf k} = U_{\bf k}+n_{\bf k}E_{\bf k} \label{vek}
\end{equation}
and
\begin{equation}
S_{\bf k} = -k_B\left[n_{\bf k}\ln n_{\bf k} - (1+n_{\bf k})\ln (1+n_{\bf k})\right], \label{Sk}
\end{equation}
respectively.\cite{fetter,huang,feynmanBook} However, because of the $T$-dependence of ${\hat H}$, when we calculate,
for a single-particle state of wave-vector {\bf k}, the partition function $Z_{\bf k} = \text{Tr}(e^{-{\hat H}_{\bf
k}/k_BT})$ (where ${\hat H}_{\bf k}=U_{\bf k}+E_{\bf k}\alpha^{\dagger}_{\bf k}\alpha_{\bf k}$), free energy $F_{\bf
k} = -k_BT\ln Z_{\bf k}$, entropy $S_{\bf k} = -\partial F_{\bf k}/\partial T$ and thermal energy $\varepsilon_{\bf
k} = F_{\bf k}+TS_{\bf k}$, we find that there are additional terms involving $\partial U_{\bf k}/\partial T$ and
$\partial E_{\bf k}/\partial T$ in each of the expressions for $\varepsilon_{\bf k}$ and $S_{\bf k}$, as compared to
Eqs. (\ref{vek}) and (\ref{Sk}).  By letting the sum of the additional terms in each of the expressions for
$\varepsilon_{\bf k}$ and $S_{\bf k}$ to be zero, we arrive at
\begin{equation}
\frac{\partial U_{\bf k}}{\partial T} + \, n_{\bf k} \frac{\partial E_{\bf k}}{\partial T} = 0. \label{haoEq00}
\end{equation}

This equation represents an additional self-consistency requirement of the theory, and must be consistent with Eq.
(\ref{gap-eq}). To see that this is indeed true, we substitute Eq. (\ref{Uk}) into Eq. (\ref{haoEq00}) to obtain
\begin{equation}
|\Delta_{\bf k}|^2\frac{\partial}{\partial T}\left(\frac{1+2n_{\bf k}}{E_{\bf k}}\right) = 0. \label{haoEq01}
\end{equation}
A solution of this equation is certainly also a solution of Eq. (\ref{gap-eq11}), and therefore also a solution of
Eq. (\ref{gap-eq}), and thus, we see that Eq. (\ref{haoEq00}) is indeed consistent with Eq. (\ref{gap-eq}).

Equation (\ref{haoEq01}) shows that there are two possible solutions for $|\Delta_{\bf k}|$ for each single-particle
state of wave-vector {\bf k}: one is a trivial solution, $|\Delta_{\bf k}| = 0$ (corresponding to the normal state),
and the other is a non-trivial solution ($|\Delta_{\bf k}|>0$, corresponding to the superfluid state) satisfying
\begin{equation}
\frac{\partial}{\partial T}\left(\frac{1+2n_{\bf k}}{E_{\bf k}}\right) = 0, \label{haoEq02}
\end{equation}
which is readily solved to give
\begin{equation}
\frac{1+2n_{\bf k}}{E_{\bf k}} = \text{$T$-independent for $T\leq T_c$}. \label{haoEq03}
\end{equation}

A consequence of Eq. (\ref{haoEq02}), or (\ref{haoEq03}), is that chemical potential $\mu$ in the superfluid state is
$T$-independent. This is shown in Appendix \ref{appMu} (where we discuss chemical potential $\mu$ and
number-of-particle distribution $\langle a^{\dagger}_{\bf k}a_{\bf k}\rangle$ in the superfluid state). Then, by
using $1+2n_{\bf k} = \coth(E_{\bf k}/2k_BT)$ and the condition that $|\Delta_{\bf k}| = 0$ at $T=T_c$, we can
express Eq. (\ref{haoEq03}) as
\begin{equation}
\frac{\coth\left(E_{\bf k}/2k_BT\right)}{E_{\bf k}}=\frac{\coth\left(\xi_{\bf k}/2k_BT_c\right)}{\xi_{\bf k}}
\label{haoEq04}
\end{equation}
with
\begin{equation}
\xi_{\bf k}=\epsilon_{\bf k}-\mu_0,
\end{equation}
\begin{equation}
E_{\bf k}=\sqrt{(\epsilon_{\bf k}-\mu_0)^2-|\Delta_{\bf k}|^2},
\end{equation}
and $\mu_0$ being the value of $\mu$ at $T=T_c$.

It is shown in Appendix \ref{appU0} (where we discuss the ground state energy of the superfluid state) that $\mu_0$
must be below a certain negative value in order for the superfluid state to be energetically favorable as compared to
the normal state.

Equation (\ref{haoEq04}) is an implicit solution for $|\Delta_{\bf k}|$ (or $E_{\bf k}$) as a function of
$\epsilon_{\bf k}$ and $T$ for given $T_c$ and $\mu_0$. A complete solution for $\Delta_{\bf k}$ is therefore a
combination of the solution  of Eq. (\ref{haoEq04}) for $|\Delta_{\bf k}|/k_BT_c$ and the solutions of Eq.
(\ref{tc-eq}) for $T_c$ and $\theta_{\bf k}$.

The present analysis with respect to the determination of $\Delta_{\bf k}$ is similar to that of Ref.
\onlinecite{hao01} with respect to the determination of the energy gap parameter in the BCS theory of
superconductivity.

\begin{figure}
\includegraphics{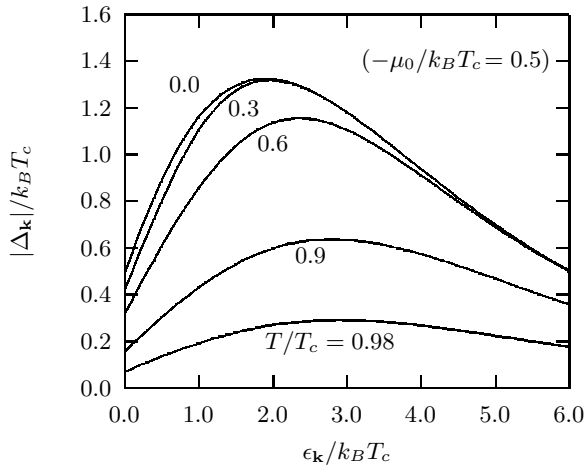}
\caption{ Energy dependence of energy gap parameter amplitude $|\Delta_{\bf k}|$ for $-\mu_0/k_BT_c=0.5$ and
different values of $T/T_c$ as indicated on the curves. } \label{figDe}
\end{figure}

\begin{figure}
\includegraphics{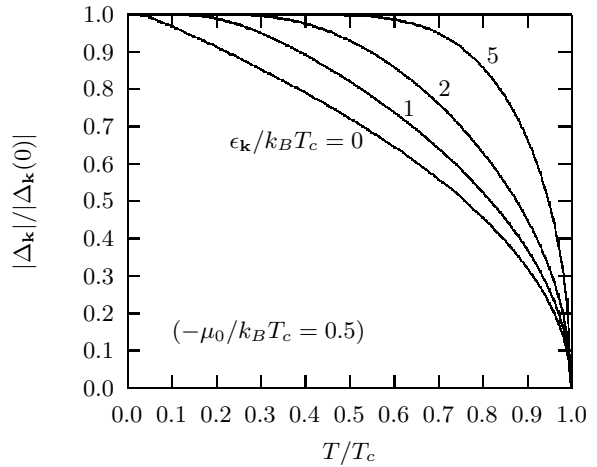}
\caption{ Temperature dependence of normalized energy gap parameter amplitude $|\Delta_{\bf k}|/|\Delta_{\bf k}(0)|$
for $-\mu_0/k_BT_c=0.5$ and different values of $\epsilon_{\bf k}/k_BT_c$ as indicated on the curves. } \label{figDt}
\end{figure}

\begin{figure}
\includegraphics{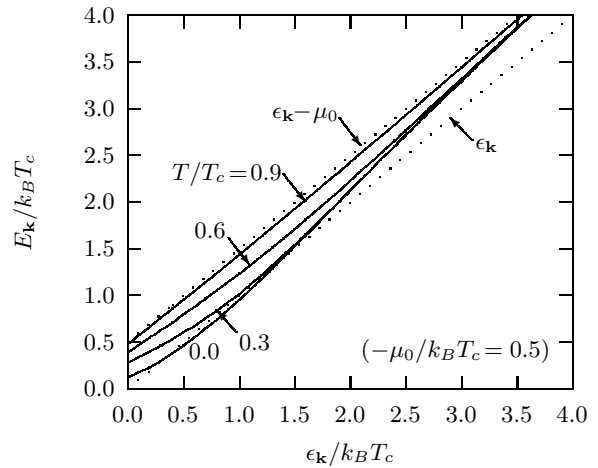}
\caption{ Superfluid-state single-particle excitation energy $E_{\bf k}$ versus normal-state single-particle energy
$\epsilon_{\bf k}$ for $-\mu_0/k_BT_c=0.5$ and different values of $T/T_c$ as indicated on the curves (solid curves).
The two dotted curves respectively show normal-state single-particle excitation energy $E_{\bf k}^{(n)}=\epsilon_{\bf
k}$ for $T$ below the Bose-Einstein condensation temperature and $E_{\bf k}^{(n)}=\epsilon_{\bf k}-\mu_0$ at $T=T_c$,
as indicated on the curves. } \label{figEk}
\end{figure}

\begin{figure}
\includegraphics{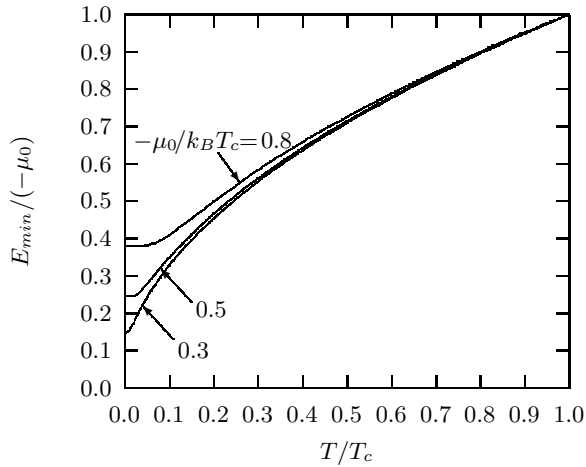}
\caption{ Temperature dependence of minimum superfluid-state single-particle excitation energy $E_{min}$ for
$-\mu_0/k_BT_c=0.3$, 0.5 and 0.8, as indicated on the curves. } \label{figEmin}
\end{figure}

We solve Eq. (\ref{haoEq04}) numerically by using an iterative method\cite{conte80} to obtain $|\Delta_{\bf
k}|/k_BT_c$ as a function of $\epsilon_{\bf k}/k_BT_c$ and $T/T_c$ for given $\mu_0/k_BT_c$.

Figure \ref{figDe} shows $\epsilon_{\bf k}$-dependence of $|\Delta_{\bf k}|$ for $-\mu_0/k_BT_c=0.5$ and different
values of $T$ as indicated on the curves.

Figure \ref{figDt} shows $T$-dependence of $|\Delta_{\bf k}|$ for $-\mu_0/k_BT_c=0.5$ and different values of
$\epsilon_{\bf k}$ as indicated on the curves.

Figure \ref{figEk} shows $E_{\bf k}$ versus $\epsilon_{\bf k}$ for $-\mu_0/k_BT_c=0.5$ and different values of $T$ as
indicated on the curves. For comparison, normal-state single-particle excitation energy $E_{\bf
k}^{(n)}=\epsilon_{\bf k}$ for $T$ below the Bose-Einstein condensation temperature\cite{fetter,huang,feynmanBook}
and $E_{\bf k}^{(n)}=\epsilon_{\bf k}-\mu_0$ at $T=T_c$ are shown as the dotted curves in the figure.

From Fig. 3 we can see the existence of an energy gap in the superfluid-state single-particle excitation spectrum.
Namely, minimum value $E_{min}$ of $E_{\bf k}$, which is located at $\epsilon_{\bf k}=0$, is greater than zero.
According to Eq. (\ref{haoEq04}), $E_{min}=(-\mu_0)\tanh\left[(-\mu_0)/2k_BT_c\right]$ at $T=0$, and increases
monotonically to $E_{min}=-\mu_0$ at $T=T_c$.  Since, as shown in Appendix \ref{appU0}, $\mu_0$ must be below a
certain negative value in the superfluid state, we see that $E_{min}>0$ in the superfluid state.

Figure \ref{figEmin} shows $T$-dependence of $E_{min}$ for different values $-\mu_0/k_BT_c$ as indicated on the
curves.

\subsection{ Specific heat $C(T)$ }
\label{subsecC}

Having obtained $|\Delta_{\bf k}(T)|$, we can calculated thermodynamic quantities of the superfluid state.  We
consider the ground state energy of the superfluid in Appendix \ref{appU0}.  We calculate in this subsection the
specific heat of the superfluid.

In the superfluid state, specific heat $C$ is given by
\begin{eqnarray}
\!\!\!\!\!\!C\!&\!\!=\!\!&\frac{\partial\langle\hat{H}\rangle}{\partial T} =
\sum_{\bf k}E_{\bf k}\frac{\partial n_{\bf k}}{\partial T} \nonumber \\
\!\!\!\!\!\!\!\!&\!\!=\!\!&\!k_B\!\!\sum_{\bf k}\!\!\frac{E^2_{\bf k}/T^2}{e^{E_{\bf k}/T}(1\!\!-\!\!e^{-E_{\bf
k}/T})^2\!\!+\!\!2(\xi_{\bf k}/T)\tanh(\xi_{\bf k}/2)},
\end{eqnarray}
where we have used Eq. (\ref{haoEq04}), and have adopted a set of dimensionless units for the last expression, in
which energies are measured in units of $k_BT_c$ and temperature in units of $T_c$.

\begin{figure}
\includegraphics{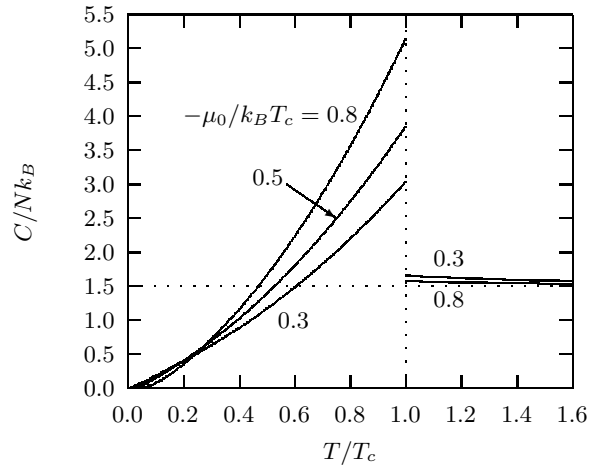}
\caption{Temperature dependence of specific heat $C$ for $-\mu_0/k_BT_c=0.3$, 0.5 and 0.8, as indicated on the
curves.  For clarity, the $T/T_c>1$ part of the $-\mu_0/k_BT_c=0.5$ curve is not shown. The horizontal dotted line
shows the value of $C/Nk_B=3/2$ for $T/T_c\gg 1$. } \label{figC}
\end{figure}

In the normal state ($T/T_c\ge 1$), the specific heat is given by
\begin{eqnarray}
C&=&\sum_{\bf k}\xi_{\bf k}\frac{\partial n_{\bf k}}{\partial T} \nonumber \\
&=&k_B\sum_{\bf k}\frac{e^{\xi_{\bf k}/T}}{(e^{\xi_{\bf k}/T}-1)^2}\left[\frac{\xi_{\bf k}^2}{T^2}-\frac{\xi_{\bf
k}}{T}\frac{\partial (-\mu)}{\partial T}\right],
\end{eqnarray}
where we have used the above-mentioned dimensionless units for the last expression.

Figure \ref{figC} shows $C/Nk_B$ versus $T/T_c$ for different values of $-\mu_0/k_BT_c$ as indicated on the curves,
where $N$ is the total number of particles of the system, and is given by
\begin{eqnarray}
N&=&\sum_{\bf k}n_{\bf k}(T_c) \nonumber \\
&=&\sum_{\bf k}\frac{1}{e^{\,\xi_{\bf k}}-1},
\end{eqnarray}
where $n_{\bf k}(T_c)$ is the Bose function at $T=T_c$, and we have used the above-mentioned dimensionless units for
the last expression.

In calculating $C/Nk_B$, we have assumed $\epsilon_{\bf k}=\hbar^2k^2/2m$ (where $m$ is particle mass), and have made
the substitution $\sum_{\bf k}\rightarrow (2\pi)^{-3}\int d^3k$.  The integrals involved are calculated by using the
Simpson method.\cite{conte80}  The method for calculating $\mu$ and $\partial\mu/\partial T$ for the specific heat in
the normal state is explained in Appendix \ref{appMu}.

As shown in Fig. \ref{figC}, there exists a finite jump in the specific heat at the transition temperature,
indicating a second-order phase transition.  The magnitude of the jump is larger for a larger value of
$-\mu_0/k_BT_c$. In the limit of $T\rightarrow 0$, we have $\partial C/\partial T\rightarrow 0$, because of the
existence of an energy gap in the single-particle excitation spectrum.

Experimentally, the $C$-versus-$T$ curve shows a $\lambda$-shaped peak at the transition.\cite{expC}

\section{ superfluidity }
\label{secSuperfluidity}

We next consider the case where the superfluid is in a state of uniform flow with velocity ${\bf v}_s$.

We write the Hamiltonian of the system as
\begin{equation}
\hat{H}=\sum_{\bf k}\left(\epsilon_{\bf k}-\mu\right)a^{\dagger}_{\bf k}a_{\bf k}+\frac{1}{2}\sum_{\bf kk'}V_{\bf
kk'}a^{\dagger}_{\bf k}a^{\dagger}_{\bf -k}a_{-\bf k'}a_{\bf k'} , \label{qH0}
\end{equation}
which is the same as the Hamiltonian of Eq. (\ref{H0}) for the case of ${\bf v}_s=0$, except that wave-vector {\bf k}
in the above expression is now measured in the coordinate frame moving with the superfluid.

We assume that pairing occurs between particles of opposite momenta measured in the coordinate frame moving with the
superfluid. I.e., we assume
\begin{equation}
\langle a_{-\bf k}a_{\bf k} \rangle \neq 0  \label{qPairing}
\end{equation}
in the superfluid state for a pair of ({\bf k}) and $(-{\bf k})$ particles.

Note that, since wave-vector {\bf k} is measured in the coordinate frame moving with the superfluid, if we use a free
Bose gas as an example, a single-particle state of wave-vector {\bf k} means, in the laboratory frame, a
single-particle state of wave function
\begin{equation}
\phi_{\bf k}=e^{i({\bf k}+{\bf q})\cdot{\bf x}} \label{phikx}
\end{equation}
and energy
\begin{equation}
\epsilon_{\bf k}=\hbar^2({\bf k}+{\bf q})^2/2m, \label{q-ek}
\end{equation}
where
\begin{equation}
{\bf q} =m{\bf v}_s/\hbar .
\end{equation}
Therefore, a pair of ({\bf k}) and $(-{\bf k})$ particles have zero net momentum in the frame moving with the
superfluid, but have a net momentum of $2\hbar{\bf q}=2m{\bf v}_s$ in the laboratory frame.

Diagonalization of Hamiltonian $\hat{H}$ is the same as in the case of ${\bf v}_s=0$, except that we now have
$\epsilon_{-\bf k}\neq\epsilon_{\bf k}$ for ${\bf v}_s\neq 0$.  The results of the diagonalization are as follows.

The diagonalized Hamiltonian is
\begin{equation}
{\hat H}=\sum_{\bf k}\left[U_{\bf k}+\frac{1}{2}\left(E_{\bf k}\alpha^{\dagger}_{\bf k}\alpha_{\bf k}+E_{-\bf
k}\alpha^{\dagger}_{-\bf k}\alpha_{-\bf k}\right)\right], \label{qH}
\end{equation}
where
\begin{equation}
U_{\bf k} = -\frac{\xi_{\bf k}+\xi_{-\bf k}}{4}+\frac{E^{(s)}_{\bf k}}{2}+\frac{|\Delta_{\bf k}|^2}{4E^{(s)}_{\bf
k}}\left(1+n_{\bf k}+n_{-\bf k}\right); \label{qUk}
\end{equation}
\begin{equation}
\xi_{\bf k} = \epsilon_{\bf k} - \mu ; \label{qxik}
\end{equation}
\begin{equation}
E_{\bf k}=E^{(s)}_{\bf k} + \frac{\xi_{\bf k}-\xi_{-\bf k}}{2}; \label{qEk}
\end{equation}
\begin{equation}
E^{(s)}_{\bf k}=\sqrt{\left(\frac{\xi_{\bf k}+\xi_{-\bf k}}{2}\right)^2-|\Delta_{\bf k}|^2 } \label{qEks}
\end{equation}
is the symmetric part of $E_{\bf k}$; and
\begin{equation}
n_{\bf k} =  \left(e^{E_{\bf k}/k_BT}-1 \right)^{-1} \label{bose}
\end{equation}
the Bose function.

Coefficients $u_{\bf k}$ and $v_{\bf k}$ are found to satisfy the following relations:
\begin{equation}
|u_{\bf k}|^2 = \frac{1}{2}\left(\frac{\xi_{\bf k}+\xi_{-\bf k}}{2E^{(s)}_{\bf k}}+1\right),
\end{equation}
\begin{equation}
|v_{\bf k}|^2 = \frac{1}{2}\left(\frac{\xi_{\bf k}+\xi_{-\bf k}}{2E^{(s)}_{\bf k}}-1\right),
\end{equation}
and
\begin{equation}
\Delta_{\bf k}u_{\bf k}v^{\star}_{\bf k} = \frac{ |\Delta_{\bf k}|^2 } { 2 E^{(s)}_{\bf k} }.
\end{equation}

The self-consistency equation for the energy gap parameter is
\begin{equation}
\Delta_{\bf k} = \mbox{} - \sum_{\bf k'} V_{\bf k,k'} \frac{1+n_{\bf k'}+n_{-\bf k'}} { 2 E^{(s)}_{\bf k'} }
\Delta_{\bf k'}. \label{q-gap-eq}
\end{equation}

Since $\epsilon_{-\bf k}\neq\epsilon_{\bf k}$ for ${\bf v}_s\neq 0$, we have $\xi_{-\bf k}\neq\xi_{\bf k}$, $E_{-\bf
k}\neq E_{\bf k}$ and $n_{-\bf k}\neq n_{\bf k}$, but we have $U_{-\bf k}=U_{\bf k}$, $E^{(s)}_{-\bf k}=E^{(s)}_{\bf
k}$ and $\Delta_{-\bf k}=\Delta_{\bf k}$, as we can see from the expressions shown above.

Chemical potential $\mu$, relative to which energies such as $\xi_{\bf k}$ and $E_{\bf k}$ are measured, is {\bf
q}-dependent, because each pair of particles in the superfluid state has a net energy increase of $2(\hbar^2q^2/2m)$
due to the flow of paired particles. Namely, we have
\begin{equation}
\mu=\mu_0+\hbar^2q^2/2m, \label{q-mu}
\end{equation}
where $\mu_0$ is the superfluid-state chemical potential for ${\bf q}=0$, and the second term is the per-particle
energy increase due to the flow of paired particles [see Appendix \ref{appMu} for a detailed derivation of Eq.
(\ref{q-mu})].

For simplicity in presenting the theory, we will use the normal-state single-particle energy spectrum of a free Bose
gas as given by Eq. (\ref{q-ek}) in the following.

With the help of Eqs. ({\ref{q-ek}) and (\ref{q-mu}), we have
\begin{equation}
\frac{\xi_{\bf k}+\xi_{-\bf k}}{2}=\xi^{(0)}_{\bf k} \label{evenXi}
\end{equation}
and
\begin{equation}
\frac{\xi_{\bf k}-\xi_{-\bf k}}{2}=\frac{\hbar^2}{m}{\bf k}\cdot{\bf q}\,, \label{oddXi}
\end{equation}
where
\begin{equation}
\xi^{(0)}_{\bf k} = \epsilon^{(0)}_{\bf k}-\mu_0
\end{equation}
and $\epsilon^{(0)}_{\bf k}$ is the value of $\epsilon_{\bf k}$ for ${\bf q}=0$.

\subsection{ $|\Delta_{\bf k}(T,{\bf q})|$ }
\label{subsecD}

The following equation is derived as an additional self-consistency requirement of the theory:
\begin{equation}
\frac{1+n_{\bf k}+n_{-\bf k}}{E^{(s)}_{\bf k}} = \text{independent of $T$ and {\bf q}}, \label{qHaoEq03}
\end{equation}
which is a generalization of Eq. (\ref{haoEq03}) to the case of ${\bf v}_s\neq 0$.  We present the details of the
derivation of this equation in Appendix \ref{appHaoEq}.

With the help of Eqs. (\ref{evenXi}) and (\ref{oddXi}) and by using $1+2n_{\bf k} = \coth(E_{\bf k}/2k_BT)$ and the
condition that $|\Delta_{\bf k}| = 0$ at $(T, {\bf q})=(T_c, 0)$, we can express Eq. (\ref{qHaoEq03}) as
\[
\frac{\coth\left(\frac{E^{(s)}_{\bf k}+2\sqrt{\epsilon^{(0)}_{\bf k}}qz_{\bf k}}{2T}\right) +
\coth\left(\frac{E^{(s)}_{\bf k}-2\sqrt{\epsilon^{(0)}_{\bf k}}qz_{\bf k}}{2T}\right)}{2E^{(s)}_{\bf k}}
\]
\begin{equation}
=\frac{\coth\left(\frac{\epsilon^{(0)}_{\bf k}-\mu_0}{2}\right)}{\epsilon^{(0)}_{\bf k}-\mu_0} \label{qHaoEq04}
\end{equation}
where
\begin{equation}
E^{(s)}_{\bf k}=\sqrt{(\epsilon^{(0)}_{\bf k}-\mu_0)^2-|\Delta_{\bf k}|^2}, \label{eqEs}
\end{equation}
$z_{\bf k}=\cos\alpha_{\bf k}$ with $\alpha_{\bf k}$ being the angle between {\bf k} and {\bf q}, and we have used a
set of dimensionless units in which energies such as $E^{(s)}_{\bf k}$, $\epsilon^{(0)}_{\bf k}$ and $\mu_0$ are
measured in units of $k_BT_c$, temperature is measured in units of $T_c$, and $q$ in units of $q_0$, which is defined
via $\hbar^2q_0^2/2m=k_BT_c$. For $^4$He, $q_0$ is $\sim 1\text{ \AA}^{-1}$, and the corresponding superfluid
velocity is $v_{s0}=\hbar q_0/m\sim 10^2$ cm/s.

\begin{figure}
\includegraphics{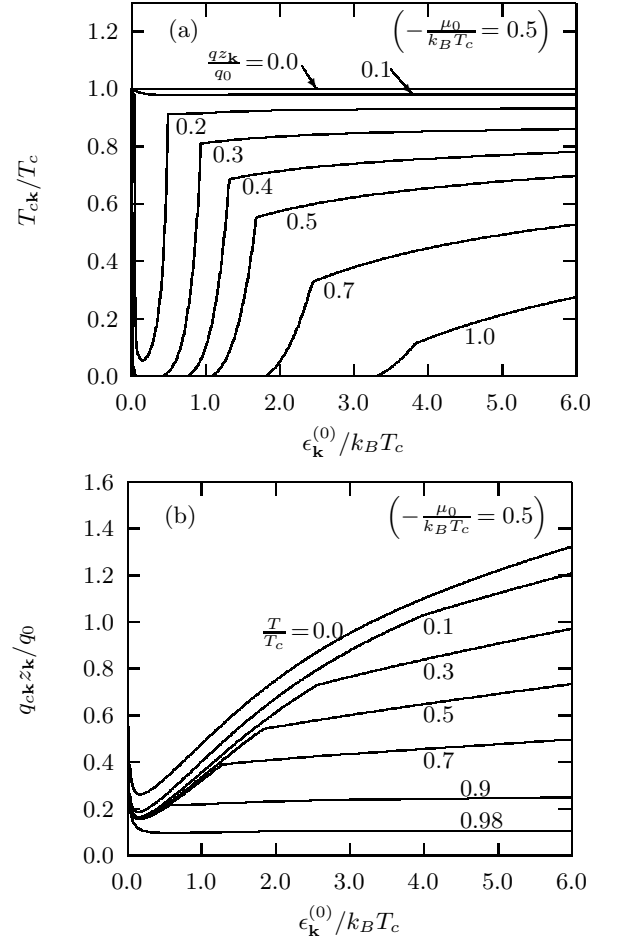}
\caption{ (a) $T_{c{\bf k}}$ versus $\epsilon^{(0)}_{\bf k}$ for different values of $qz_{\bf k}/q_0$ as indicated on
the curves; (b) $q_{c{\bf k}}z_{\bf k}$ versus $\epsilon^{(0)}_{\bf k}$ for different values of $T/T_c$ as indicated
on the curves; $-\mu_0/k_BT_c=0.5$ in both (a) and (b). } \label{figTckQck}
\end{figure}

\begin{figure}
\includegraphics{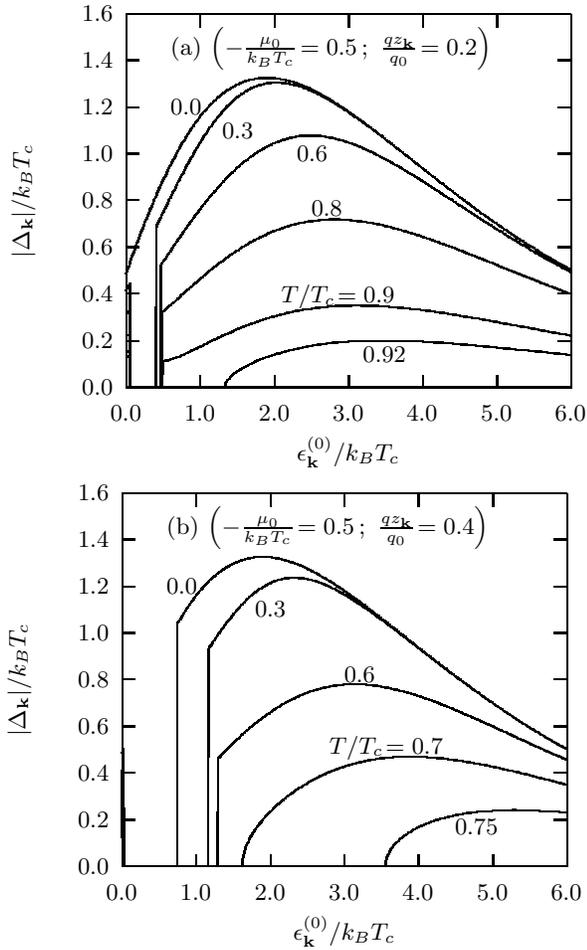}
\caption{ $|\Delta_{\bf k}|$ versus $\epsilon^{(0)}_{\bf k}$ for different values of $T$ as indicated on the curves;
$qz_{\bf k}/q_0=0.2$ and $0.4$ in (a) and (b), respectively; $-\mu_0/k_BT_c=0.5$ in both (a) and (b). }
\label{figDeQ}
\end{figure}

\begin{figure}
\includegraphics{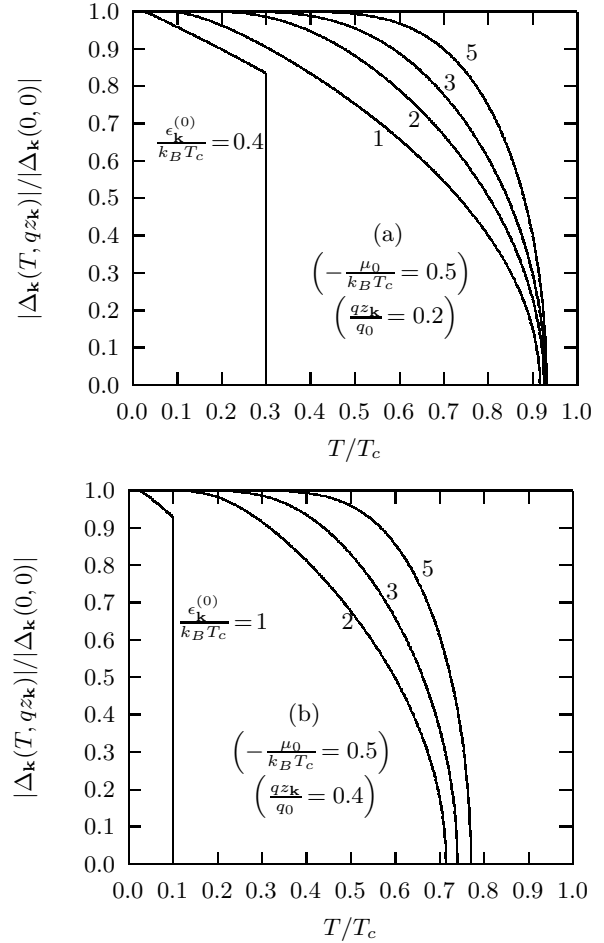}
\caption{ $|\Delta_{\bf k}(T,qz_{\bf k})|/|\Delta_{\bf k}(0,0)|$ versus $T/T_c$ for different values of
$\epsilon^{(0)}_{\bf k}/k_BT_c$ as indicated on the curves; $qz_{\bf k}/q_0=0.2$ and $0.4$ in (a) and (b),
respectively; $-\mu_0/k_BT_c=0.5$ in both (a) and (b). } \label{figDtQ}
\end{figure}

\begin{figure}
\includegraphics{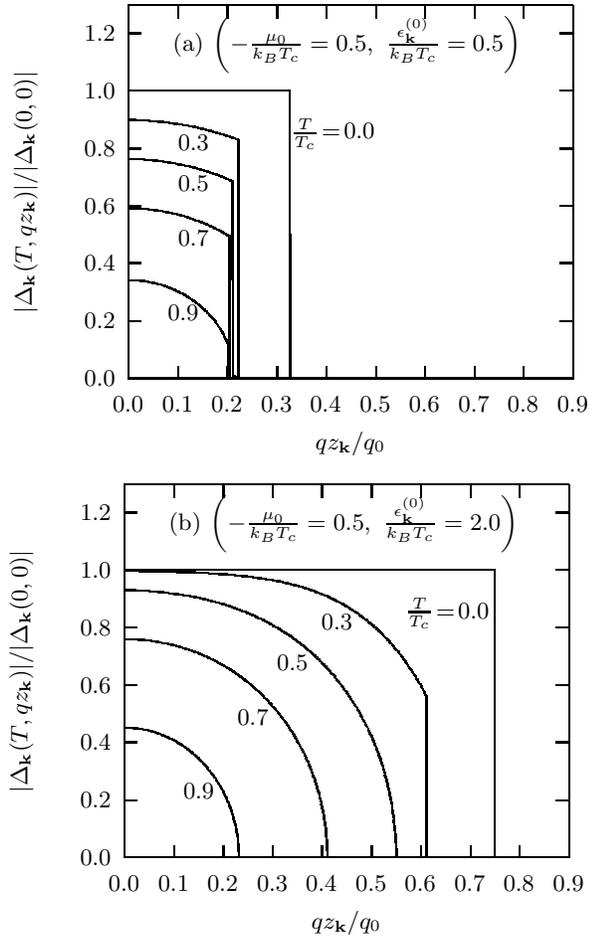}
\caption{ $|\Delta_{\bf k}(T,qz_{\bf k})|/|\Delta_{\bf k}(0,0)|$ versus $qz_{\bf k}/q_0$ for different values of
$T/T_c$ as indicated on the curves; $\epsilon^{(0)}_{\bf k}/k_BT_c=0.5$ and $2.0$ in (a) and (b), respectively;
$-\mu_0/k_BT_c=0.5$ in both (a) and (b). } \label{figDq}
\end{figure}

Equation (\ref{qHaoEq04}) is an implicit solution for $|\Delta_{\bf k}(T,{\bf q})|$ [or $E^{(s)}_{\bf k}(T,{\bf
q})$]. The variables $({\bf k}, T, {\bf q})$ for the function $|\Delta_{\bf k}(T,{\bf q})|$ appear in Eq.
(\ref{qHaoEq04}) in the forms of $(\epsilon^{(0)}_{\bf k}, T, qz_{\bf k})$, where $qz_{\bf k}$ is the component of
{\bf q} along {\bf k}. We solve Eq. (\ref{qHaoEq04}) by using an iterative method\cite{conte80} to obtain
$|\Delta_{\bf k}|$ as a function of $\epsilon^{(0)}_{\bf k}$, $T$ and $qz_{\bf k}$ for given $\mu_0$.

Note that temperature $T_{c\bf k}$, defined as such that Eq. (\ref{qHaoEq04}) has no $|\Delta_{\bf k}|>0$ solution
for given $\epsilon^{(0)}_{\bf k}$, $T$ and $qz_{\bf k}$ if $T\ge T_{c\bf k}$, is a function of $\epsilon^{(0)}_{\bf
k}$ and $qz_{\bf k}$. Only for ${\bf q}=0$ is $T_{c\bf k}=T_c$ the same for all single-particle states.

Similarly, superfluid wave-vector $q_{c\bf k}$, defined as such that Eq. (\ref{qHaoEq04}) has no $|\Delta_{\bf k}|>0$
solution for given $\epsilon^{(0)}_{\bf k}$, $T$ and $qz_{\bf k}$ if $q\ge q_{c\bf k}$, is a function of
$\epsilon^{(0)}_{\bf k}$, $T$ and $z_{\bf k}$ (or, $q_{c\bf k}z_{\bf k}$ is a function of $\epsilon^{(0)}_{\bf k}$
and $T$).

Figure \ref{figTckQck}(a) shows $T_{c\bf k}$ versus $\epsilon^{(0)}_{\bf k}$ for different values of $qz_{\bf k}$,
and Fig. \ref{figTckQck}(b) shows $q_{c{\bf k}}z_{\bf k}$ versus $\epsilon^{(0)}_{\bf k}$ for different values of
$T$; $-\mu_0/k_BT_c=0.5$ is assumed in both Fig. \ref{figTckQck}(a) and Fig. \ref{figTckQck}(b).

Note that there is an upper bound for $T_{c\bf k}$, i.e., $T_{c\bf k}\le T_c$, but there is no upper bound for
$q_{c\bf k}$, i.e., $q_{c\bf k}\rightarrow\infty$ for $\epsilon^{(0)}_{\bf k}\rightarrow\infty$, and $q_{c\bf
k}=\infty$ if ${\bf k}\perp{\bf q}$ (because $z_{\bf k}=0$ for this case).

The minimum values of $T_{c\bf k}$ and $q_{c\bf k}$,  $T_{{c\bf k},\text{min}}(qz_{\bf k})$ and $q_{{c\bf
k},\text{min}}(T)$, are of particular importance.  For $T<T_{{c\bf k},\text{min}}(qz_{\bf k})$ [or $q<q_{{c\bf
k},\text{min}}(T)$], the system is in an all-paired state, in which $|\Delta_{\bf k}|>0$ for all particles.  For
$T_{{c\bf k},\text{min}}(qz_{\bf k})<T<T_c$ [or $q_{{c\bf k},\text{min}}(T)<q$], the system is in a partly-paired
state, in which particles in states having $T_{c\bf k}(qz_{\bf k})<T$ [or $q_{c\bf k}(T)<q$] become de-paired (having
$|\Delta_{\bf k}|=0$) while those in states having $T_{c\bf k}(qz_{\bf k})>T$ [or $q_{c\bf k}(T)>q$] remain paired
(having $|\Delta_{\bf k}|>0$).  For $T>T_c$, the system is in the normal state, in which $|\Delta_{\bf k}|=0$ for all
particles.

A finite viscosity should be observable in a partly-paired state, because of the existence of de-paired particles,
which are expected to behave as normal-state particles. We therefore expect critical velocity $v_{sc}(T)$, defined as
the superfluid velocity at the onset of an observable viscosity, to be about the same as $\hbar q_{{c\bf
k},\text{min}}(T)/m$.  From the numerical results shown in Fig. \ref{figTckQck}(b), for example, which are obtained
for the case of $-\mu_0/k_BT_c=0.5$, we can see that $q_{{c\bf k},\text{min}}(T)$ is a few tenth of $q_0$,
corresponding to a superfluid velocity of a few tenth of $v_{s0}=\hbar q_0/m$.  Since $v_{s0}\sim 10^2$ cm/s for
$^4$He, we see that the critical velocity for $^4$He in this case is about a few tens of centimeters per second
(which is several orders of magnitude smaller than the value previously predicted by Landau\cite{landau}).

Numerical results for $|\Delta_{\bf k}|$ versus $\epsilon^{(0)}_{\bf k}$ for different values of $T$ and $qz_{\bf k}$
are shown in Fig. \ref{figDeQ} [Figs. \ref{figDeQ}(a) and \ref{figDeQ}(b)]. Figure \ref{figDeQ}(a) shows an example
of the case of $0<q<q_{c{\bf k},\text{min}}(0)$. In this case, the $|\Delta_{\bf k}|$-versus-$\epsilon^{(0)}_{\bf k}$
curve for $T=0$ is the same as in the case of ${\bf q}=0$ (which is shown in Fig. \ref{figDe}). As $T$ increases,
$|\Delta_{\bf k}|$ for smaller $\epsilon^{(0)}_{\bf k}$ is more strongly suppressed, and decreases faster.  As $T$
increases further so that $T>T_{c{\bf k},\text{min}}$, the $|\Delta_{\bf k}|$-versus-$\epsilon^{(0)}_{\bf k}$ curve
has a $|\Delta_{\bf k}|=0$ part for low energies (except for $\epsilon^{(0)}_{\bf k}\rightarrow 0$), for which
$T_{c{\bf k}}(qz_{\bf k})<T$ [or $q_{c{\bf k}}(T)<q$]. Figure \ref{figDeQ}(b) shows an example of the case of
$q>q_{c{\bf k},\text{min}}(0)$. In this case, the $|\Delta_{\bf k}|$-versus-$\epsilon^{(0)}_{\bf k}$ curve has a
$|\Delta_{\bf k}|=0$ part even at $T=0$. Namely, at $T=0$, $|\Delta_{\bf k}|=0$ for those single-particle states with
$q_{c{\bf k}}(0)<q$.  The vertical rises (or drops) in the $|\Delta_{\bf k}|$-versus-$\epsilon^{(0)}_{\bf k}$ curves
in both Fig. \ref{figDeQ}(a) and Fig. \ref{figDeQ}(b) indicate discontinuities.

Figures \ref{figDtQ} and \ref{figDq} show, respectively, the numerical results for the $T$-dependence and $qz_{\bf
k}$-dependence of $|\Delta_{\bf k}|$. As shown in the figures, $|\Delta_{\bf k}|$ is a monotonic decreasing function
of $T$ and $qz_{\bf k}$, except that, at $T=0$, $|\Delta_{\bf k}|>0$ is a constant for $q<q_{c{\bf k}}(0)$ (Fig.
\ref{figDq}). The vertical drops in some of the curves shown in Figs. \ref{figDtQ} and \ref{figDq} indicate
discontinuities.

\subsection{ Superfluid particle density $n_s$ }

Particle current density {\bf j} is the expectation value of particle current density operator $\hat{\bf
J}$.\cite{fetter} I.e.,
\begin{equation}
{\bf j} = \langle\hat{\bf J}\rangle,
\end{equation}
where
\begin{equation}
\hat{\bf J} = \frac{1}{2}\left[\hat{\Psi}^\dagger\left(\hat{\bf v}\hat{\Psi}\right) + \left(\hat{\bf
v}\hat{\Psi}\right)^\dagger\hat{\Psi}\right],
\end{equation}
the particle field operator
\begin{equation}
\hat{\Psi}=\sum_{\bf k}\phi_{\bf k}({\bf x})a_{\bf k}
\end{equation}
with $\phi_{\bf k}({\bf x})$ being the single-particle wave-function [given by Eq. (\ref{phikx})], and the velocity
operator
\begin{equation}
\hat{\bf v}= -i\frac{\hbar}{m}\nabla.
\end{equation}

A straightforward calculation gives
\begin{equation}
{\bf j}=n\frac{\hbar}{m}\,{\bf q} -\frac{\hbar}{2m}\sum_{\bf k}\left(n_{-\bf k}-n_{\bf k}\right){\bf k}, \label{eqJ1}
\end{equation}
where $n$ is the particle density and can be expressed as (see Appendix \ref{appMu})
\begin{equation}
n=\sum_{\bf k}n_{\bf k}(T_c),
\end{equation}
\begin{equation}
n_{\pm\bf k}=\left[e^{\left(\sqrt{(\epsilon^{(0)}_{\bf k}-\mu_0)^2-|\Delta_{\bf k}|^2}\pm\frac{\hbar^2}{m}{\bf
k}\cdot{\bf q}\right)/k_BT}-1\right]^{-1},
\end{equation}
and $n_{\bf k}(T_c)$ is the value of $n_{\bf k}$ at $(T,{\bf q})=(T_c,0)$.

The first term on the right-hand side of Eq. (\ref{eqJ1}) represents a uniform flow of all particles.  The second
term represents contribution from single-particle excitations and de-paired particles, and tends to cancel the first
term. When all particles are in the superfluid ground state (at $T=0$ and for $q$ below a threshold), the second term
is zero. On the other hand, when $|\Delta_{\bf k}|=0$ for all single-particle states, the two terms cancel each other
so that ${\bf j}=0$.

Since superfluid current density ${\bf j}=0$ without pairing, as shown above, according to standard quantum theory of
many-particle systems, it is clear that superfluidity arises from pairing of particles, not from Bose-Einstein
condensation (there is no microscopic theoretical justification for the view that Bose-Einstein condensation leads to
superfluidity). Therefore, we believe that the superfluid properties of liquid $^4$He,\cite{londonII,hess-fairbank}
as well as the recently observed superfluid properties of ultra-cold atomic gases (such as the persistent flow of
atoms in a toroidal trap\cite{phillips} and the vortices in rotating atomic gases\cite{ketterle,cornel}), are
associated with pairing of the atoms involved, not Bose-Einstein condensation.

By using $\epsilon^{(0)}_{\bf k}=\hbar^2k^2/2m$ and making the substitution $\sum_{\bf
k}\!\!\rightarrow(2\pi)^{-3}\!\int d^3k$, we can rewrite Eq. (\ref{eqJ1}) as
\begin{equation}
{\bf j} = n_s\frac{\hbar}{m}\,{\bf q} = n_s{\bf v}_s , \label{eqJ2}
\end{equation}
where $n_s$ is the effective superfluid particle density, and, by using the above-introduced dimensionless units (in
which energies are measured in units of $k_BT_c$, temperature $T$ in units of $T_c$ and superfluid wave-vector $q$ in
units of $q_0$), can be expressed as
\begin{equation}
\frac{n_s}{n} = 1 - \frac{1}{2q\tilde{n}}\int^\infty_0d\epsilon^{(0)}_{\bf k}\epsilon^{(0)}_{\bf k}\int^1_0dz_{\bf
k}z_{\bf k} \left(n_{-\bf k}-n_{\bf k}\right), \label{ns}
\end{equation}
where
\begin{equation}
\tilde{n} = \int^\infty_0\frac{d\epsilon^{(0)}_{\bf k}(\epsilon^{(0)}_{\bf k})^{1/2}}{e^{(\epsilon^{(0)}_{\bf
k}-\mu_0)}-1} \label{ntilde}
\end{equation}
is a function of $\mu_0$ and relates to $n$ (i.e., $n\propto\tilde{n}$), and
\begin{equation}
n_{\pm\bf k}=\left[e^{\left(\sqrt{(\epsilon^{(0)}_{\bf k}-\mu_0)^2-|\Delta_{\bf k}|^2}\pm 2\sqrt{\epsilon^{(0)}_{\bf
k}} qz_{\bf k}\right)/T}\!\!\!-1\right]^{-1}\!\!\!\!\!\!. \label{npmk}
\end{equation}

\subsubsection{$n_s$ for $q\rightarrow 0$}

In the limit of $q\rightarrow 0$, Eq. (\ref{ns}) becomes
\begin{equation}
\frac{n_s}{n} = 1 - \frac{2}{3\tilde{n}T}\int^\infty_0\frac{d\epsilon^{(0)}_{\bf k}(\epsilon^{(0)}_{\bf k})^{3/2}
e^{E^{(s)}_{\bf k}/T}}{\left(e^{E^{(s)}_{\bf k}/T}-1\right)^2}, \label{nsq0}
\end{equation}
where $E^{(s)}_{\bf k}$, as a function of $\epsilon^{(0)}_{\bf k}$ and $T$ for given $\mu_0$, is determined by Eq.
(\ref{qHaoEq04}) for $q=0$.

\begin{figure}
\includegraphics{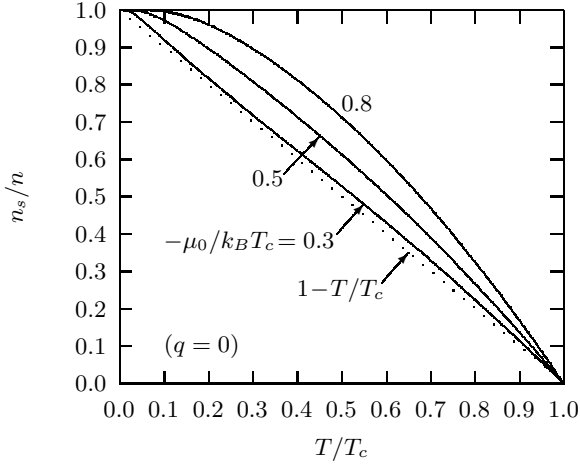}
\caption{Temperature dependence of superfluid particle density $n_s$ for the case of $q\rightarrow 0$ for
$-\mu_0/k_BT_c=0.3$, 0.5 and 0.8, as indicated on the curves. (The dotted curve shows a linear $1-T/T_c$
dependence.)} \label{figNsQ00}
\end{figure}

Numerical results for $n_s/n$ for the case of $q\rightarrow 0$ as a function of $T$ are shown in Fig. \ref{figNsQ00}
for different values of $\mu_0$. Note that $dn_s/dT\rightarrow 0$ in the limit of $T\rightarrow 0$, because of the
existence of an energy gap in the superfluid single-particle excitation spectrum.

\subsubsection{$n_s$ for finite $q$}

In Appendix \ref{appMu}, chemical potential $\mu$ in the superfluid state is determined based on the assumption that
all particles are paired in the superfluid state.  For the case of finite $q$, de-paired particles may exist.  The
result for $\mu$ obtained in Appendix \ref{appMu} is no longer valid when de-paired particles exist.  The question
how to determine the chemical potential in a partly-paired state is not addressed in this paper. To proceed, we make
the following approximation with respect to $n_{\pm\bf k}$ in calculating $n_s$ when de-paired particles exist. For
paired particles (for which $|\Delta_{\bf k}|>0$), we use the same result for $n_{\pm\bf k}$ as in the case of an
all-paired state; and for de-paired particles (for which $|\Delta_{\bf k}|=0$), we use the following approximation:
\begin{eqnarray}
n_{-\bf k} - n_{\bf k} &=& \langle a^{\dagger}_{-\bf k}a_{-\bf k}\rangle - \langle a^{\dagger}_{\bf k}a_{\bf
k}\rangle \\
&\simeq& n^{(n)}_{-\bf k}(T_c) - n^{(n)}_{\bf k}(T_c)\,,
\end{eqnarray}
where
\begin{equation}
n^{(n)}_{\pm\bf k}(T_c)=\left[e^{\left(\epsilon_{\pm\bf k}-\mu_0\right)}-1\right]^{-1}
\end{equation}
(in dimensionless units) with $\epsilon_{\pm\bf k}=\hbar^2(\pm{\bf k}+{\bf q})^2/2m$.

\begin{figure}
\includegraphics{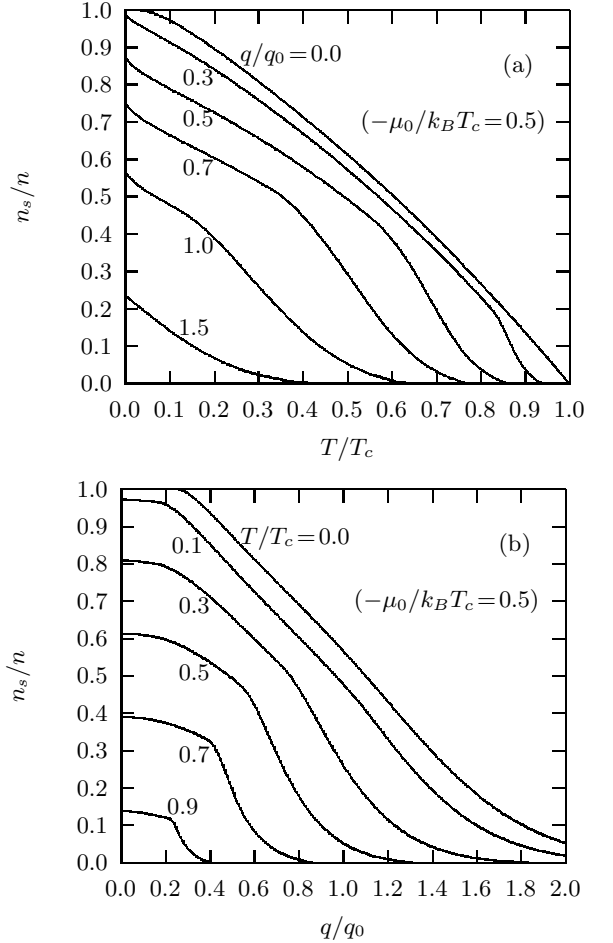}
\caption{(a) $n_s/n$ versus $T/T_c$ for different values of $q/q_0$ as indicated on the curves; (b) $n_s/n$ versus
$q/q_0$ for different values of $T/T_c$; $-\mu_0/k_BT_c=0.5$ in both (a) and (b). } \label{figNs}
\end{figure}

Numerical results for $n_s/n$ obtained based on the above-described approximation are shown in Fig. \ref{figNs}(a) as
$n_s/n$ versus $T/T_c$ for difference values of $q/q_0$ and in Fig. \ref{figNs}(b) as $n_s/n$ versus $q/q_0$ for
difference values of $T/T_c$.  In both Fig. \ref{figNs}(a) and Fig. \ref{figNs}(b), $-\mu_0/k_BT_c=0.5$ is assumed.

\subsection{ Free energy density $F$ }

From diagonalized Hamiltonian $\hat{H}$ [Eq. (\ref{qH})], we derive the following expression for the free energy
density in the superfluid state:
\begin{eqnarray}
F &\!\!\! = &\!\!\!\sum_{\bf k}\!\left[ U_{\bf k} \!+\! \frac{k_BT}{2} \ln \!\left( 1 \!-\! e^{-\frac{E_{\bf
k}}{k_BT}}
\!\right)\!\! \left( 1 \!-\! e^{-\frac{E_{-\bf k}}{k_BT}}\!\right) \!\right] \nonumber \\
& & \mbox{} + \,\,\, n\frac{\hbar^2q^2}{2m}, \label{F}
\end{eqnarray}
where the first term comes from $\,\,-k_BT\ln \left[ \text{Tr} \left( e^{-\hat{H}/k_BT} \right) \right]$, which is
the usual statistical free energy density,\cite{fetter,huang,feynmanBook} and the second term comes from $\,n(\mu
\!-\! \mu_0)$, which is the energy increase due to the flow of the superfluid, and which is added to $F$ because
single-particle energies in the expression for Hamiltonian $\hat{H}$ are measured relative to $\mu$.

For an isotropic system (such as liquid $^4$He), $F$ is a function of $T$ and $q=|{\bf q}|$, i.e., $F=F(T,q)$. As can
be shown, superfluid current density {\bf j} and effective superfluid particle density $n_s$ are related to $F$ via
the relations
\begin{equation}
{\bf j} = \frac{1}{\hbar}\frac{\partial F}{\partial {\bf q}} \label{jF}
\end{equation}
and
\begin{equation}
n_s = \frac{m}{\hbar^2q}\frac{\partial F}{\partial q}\,, \label{nsF}
\end{equation}
respectively [where $F$ is as given by Eq. (\ref{F})].  When $|\Delta_{\bf k}|=0$ for all single-particle states, as
in the normal state, $F$ becomes {\bf q}-independent, and we have ${\bf j}=0$ and $n_s=0$.

\subsection{ Spatially varying ${\bf v}_s({\bf x})$ }

The theory presented so far is based on the assumption that superfluid velocity ${\bf v}_s$ is spatially constant.
We can extend the theory to the case where superfluid velocity ${\bf v}_s$ is spatially varying, by making an
assumption that there exists a length $l$ such that the following is true: Length $l$ is large compared to
inter-particle distance so that the properties of particles in volume $l^3$ are essentially those of an infinite
system, but small by macroscopic standards so that the volume can be regarded as a ``point'' macroscopically and all
thermodynamic functions of the system vary negligibly over the distance $l$.

Based on this assumption, quantities such as $|\Delta_{\bf k}|$, $n_s$, {\bf j} and $F$ can all be considered as
local quantities, obtained with respect to particles in a volume $l^3$ around a local point {\bf x} for ${\bf q}={\bf
q}({\bf x})$, i.e., $|\Delta_{\bf k}|=|\Delta_{\bf k}(T,{\bf q}({\bf x}))|$, $n_s=n_s(T,q({\bf x}))$, ${\bf j}={\bf
j}(T,{\bf q}({\bf x}))$ and $F=F(T,q({\bf x}))$, where ${\bf q}({\bf x}) = m{\bf v}_s({\bf x})/\hbar$ is assumed to
vary spatially with a length much larger than $l$.

The theory presented so far for the case of ${\bf v}_s\neq 0$ is similar to the theory presented in Ref.
\onlinecite{hao11} for superconductivity in the presence of a magnetic field.

\subsection{ Hess-Fairbank effect }

Since $n_s>0$ in the superfluid state, we see from Eq. (\ref{nsF}) that we have
\begin{equation}
\frac{\partial F}{\partial q} > 0,
\end{equation}
which shows that a larger value of $q$ is energetically less favorable in the superfluid state.  This implies that a
superfluid tends to expel superfluid wave-vector ${\bf q}({\bf x})$, or, equivalently, superfluid velocity ${\bf
v}_s({\bf x})=\hbar{\bf q}({\bf x})/m$, from its interior so as to minimize the overall free energy of the system.
This result qualitatively explains the Hess-Fairbank effect,\cite{hess-fairbank} the reduction of moment of inertia
of a rotating cylinder of liquid $^4$He when it is cooled through the superfluid transition. Namely, when the liquid
is in the superfluid state, because a motionless state is energetically more favorable, it stops rotating with the
container, except in the immediate vicinity of the wall of the container where the liquid rotates with the container
due to interaction between the liquid and the wall at the interface.

Although the theory presented so far provides a qualitative explanation for the Hess-Fairbank effect, it does not
allow quantitative description of the Hess-Fairbank effect.  Namely, the theory says that a superfluid tends to expel
superfluid velocity ${\bf v}_s({\bf x})$, but it does not tell us how ${\bf v}_s({\bf x})$ can be determined for
given boundary condition and temperature (in the case of a rotating cylinder of liquid $^4$He, for example, the
boundary condition is determined by the angular speed and geometry of the container).

We note that the Hess-Fairbank effect is analogous to the Meissner effect in superconductors, and the latter is
quantitatively describable by combination of the London equation\cite{londonI} ${\bf j}=-n_s{\bf a}$ and the Ampere's
law ${\bf j}=\nabla\times{\bf b}$ (here {\bf j}, $n_s$, {\bf a} and ${\bf b}=\nabla\times{\bf a}$ are, respectively,
the electrical current density, superconducting electron density, vector potential and magnetic flux density, and a
set of dimensionless units is used for the present discussion).  Our Eq. (\ref{eqJ2}) is analogous to the London
equation, with ${\bf v}_s$ playing the role of {\bf a}.  What is missing for a superfluid is an equation analogous to
the Ampere's law.

We therefore speculate the existence of the following equation:
\begin{equation}
{\bf j} = -\Lambda\nabla\times{\bf\text{\boldmath $\omega$}}, \label{haoEq000}
\end{equation}
where
\begin{equation}
\text{\boldmath $\omega$}=\nabla\times{\bf v}_s \label{omega}
\end{equation}
is superfluid vorticity, and $\Lambda$ a positive constant.

Note that Eq. (\ref{haoEq000}) applies only to particles (or atoms) in the superfluid state; i.e., here {\bf j} and
{\boldmath$\omega$} are associated with the superfluid component of a fluid, and correspond, respectively, to ${\bf
j}_s$ and {\boldmath$\omega$}$_s$ in a two-fluid model\cite{londonII,london1938,tisza,landau}.

Equation (\ref{haoEq000}) is clearly only a speculation based on the similarity between superfluidity and
superconductivity, and thus, must derive its validity from experimental confirmation of the consequences that it
implies.

Contrary to the common view that superfluid is irrotational (vorticity-free) (for which there is no microscopic
theoretical justification), Eq. (\ref{haoEq000}) shows that, analogous to that electrical current creates magnetic
field, superfluid current creates vorticity.

As we will see below, an important consequence of Eq. (\ref{haoEq000}) is the existence of a penetration depth that
characterizes the typical distance to which superfluid velocity and superfluid vorticity penetrate into a superfluid.
This penetration depth is analogous to the London penetration depth\cite{londonI} that characterizes the typical
distance to which magnetic vector potential and magnetic field penetrate into a superconductor.

We further speculate the existence of an additional term in the expression for free energy density $F$, i.e.,
\begin{eqnarray}
F &\!\!\! = &\!\!\!\sum_{\bf k}\!\left[ U_{\bf k} \!+\! \frac{k_BT}{2} \ln \!\left( 1 \!-\! e^{-\frac{E_{\bf
k}}{k_BT}} \!\right)\!\! \left( 1 \!-\! e^{-\frac{E_{-\bf k}}{k_BT}}\!\right) \!\right] \nonumber \\
& & \mbox{} + \,\,\, n\frac{\hbar^2q^2}{2m} + \frac{\Lambda m}{2}\omega^2, \label{F2}
\end{eqnarray}
where the first two terms are the same as in Eq. (\ref{F}), and the third term, which is analogous to the magnetic
filed energy density in the case of superconductors, is the additional term whose existence is speculated. The reason
for this speculation is as follows. We note that Eq. (\ref{eqJ2}) is an equilibrium property of a superfluid, and
thus, must also be derivable as a result of the variational problem that, in the thermodynamic equilibrium, the
overall free energy of the superfluid, given by the volume integral of free energy density $F$, is stationary with
respect to arbitrary variation of ${\bf v}_s({\bf x})$.  This is true when $F$ is as given by Eq. (\ref{F2}), as can
be shown with the help of Eq. (\ref{haoEq000}).

Combination of Eqs. (\ref{eqJ2}) and (\ref{haoEq000}) allows quantitative description of the Hess-Fairbank effect.
In the following we present two simple examples.

\subsubsection{Superfluid in a rotating cylinder}

We consider a superfluid in a rotating cylinder.  Assuming the length of the cylinder is much larger than its radius
$R$, and neglecting the bottom portion of the cylinder, in terms of cylindrical coordinates $(r, \phi, z)$ and unit
vectors ($\hat{\bf r}$, {\boldmath$\hat\phi$}, $\hat{\bf z}$), we can write superfluid current density ${\bf
j}=j(r)${\boldmath $\hat\phi$}, superfluid velocity ${\bf v}_s=v_s(r)${\boldmath $\hat\phi$} and superfluid vorticity
{\boldmath $\omega$}$=\omega(r)\hat{\bf z}$, and we have
\begin{equation}
\Lambda\omega'(r)=n_s(r)v_s(r) \label{omegaprime}
\end{equation}
and
\begin{equation}
v_s'(r) = \omega(r) - \frac{v_s(r)}{r}, \label{vsprime}
\end{equation}
where a ``prime'' indicates a derivative with respect to $r$, $n_s(r)=n_s(v_s(r))$ is given by Eq. (\ref{ns}), the
first equation comes from combining Eqs. (\ref{eqJ2}) and (\ref{haoEq000}), and the second equation comes from Eq.
(\ref{omega}).  These equations describe only the behavior of the superfluid component of the fluid.  We will not
consider in this paper the behavior of the normal-fluid component of the fluid.

This is a second-order boundary value problem (which is expressed here as a system of two first-order differential
equations) with the boundary conditions
\begin{equation}
v_s(0)=0
\end{equation}
and
\begin{equation}
v_s(R)=R\Omega_0\,,
\end{equation}
where $R$ is the inner radius of the cylindrical container and $\Omega_0$ the angular speed of the container.  Here
we have assumed that, in equilibrium, $v_s(R)$ is the same as the linear speed of the inner wall of the container, as
otherwise there would be momentum transfer between the superfluid and the container.

In this paper, we will not attempt to solve this boundary value problem for arbitrary $T$ and $v_s(R)$.  Instead, for
simplicity in presenting the main features of the theory, we will consider only the case where $n_s$ is spatially
constant. As we can see from Fig. \ref{figNs}(b), that $n_s$ is spatially constant is true only at $T=0$ for
$v_s<\hbar q_{c{\bf k},\text{min}}(0)/m$ [i.e., $n_s$ is independent of $v_s$ at $T=0$ for $v_s<\hbar q_{c{\bf
k},\text{min}}(0)/m$], and is approximately true at higher temperatures for sufficiently low values of $v_s$.

For a spatially constant $n_s$, the above-described boundary value problem can be solved analytically, and the
solutions are:
\begin{equation}
v_s(r) = R\Omega_0\frac{I_1(r/\lambda)}{I_1(R/\lambda)}
\end{equation}
and
\begin{equation}
\omega(r) = \frac{R\Omega_0}{\lambda}\frac{I_0(r/\lambda)}{I_1(R/\lambda)},
\end{equation}
where $I_n(x)$ is the modified Bessel function of the first kind of order $n$,\cite{spiegel} and
\begin{equation}
\lambda = \sqrt{\Lambda/n_s}\,\,\,. \label{lambda}
\end{equation}

For $R/\lambda\gg 1$, by using the asymptotic expansion\cite{spiegel} $I_n(x)\sim e^x/\sqrt{2\pi x}$, where $x\gg 1$,
we have, near the inner wall of the container,
\begin{equation}
v_s(r) \simeq R\Omega_0\,e^{-(R-r)/\lambda}
\end{equation}
and
\begin{equation}
\omega(r) \simeq \frac{R\Omega_0}{\lambda}\,e^{-(R-r)/\lambda},
\end{equation}
from which we see that superfluid velocity and superfluid vorticity ``penetrate'' only a distance of the order of
$\lambda$ into the superfluid; at a depth of little more than $\lambda$, superfluid velocity and superfluid vorticity
are practically zero; and thus, $\lambda$ has the meaning of ``penetration depth'' that characterizes the distance to
which superfluid velocity and superfluid vorticity penetrate into a superfluid, analogous to the London penetration
depth\cite{londonI} that characterizes the distance to which magnetic field penetrates into a superconductor.

For $R/\lambda\ll 1$, by using the approximation\cite{spiegel} $I_0(x)\sim 1+O(x^2)$ and $I_1(x)\sim x/2 + O(x^3)$,
where $x\ll 1$, we have
\begin{equation}
v_s(r) \simeq \Omega_0r
\end{equation}
and
\begin{equation}
\omega(r) \simeq 2\Omega_0,
\end{equation}
which are the same as the results for a rotating rigid body.

\subsubsection{Flow of superfluid in a pipe}

We next consider the case where a superfluid flows through an infinitely long pipe of a constant circular cross
section. Let the axis of the pipe be the $z$-axis, the inner radius of the pipe be $R$, and the total superfluid
current be $I$. In terms of cylindrical coordinates $(r, \phi, z)$ and unit vectors ($\hat{\bf r}$,
{\boldmath$\hat\phi$}, $\hat{\bf z}$), we can write superfluid current density ${\bf j}=j(r)\hat{\bf z}$, superfluid
velocity ${\bf v}_s=v_s(r)\hat{\bf z}$ and superfluid vorticity {\boldmath
$\omega$}$=\omega(r)${\boldmath$\hat\phi$}, and we have
\begin{equation}
\omega'(r)=-\frac{\omega(r)}{r}-\frac{n_s(r)}{\Lambda}v_s(r) \label{omegaprime2}
\end{equation}
and
\begin{equation}
v_s'(r) = -\omega(r), \label{vsprime2}
\end{equation}
where a ``prime'' indicates a derivative with respect to $r$, $n_s(r)=n_s(v_s(r))$ is given by Eq. (\ref{ns}), the
first equation comes from combining Eqs. (\ref{eqJ2}) and (\ref{haoEq000}), and the second equation comes from Eq.
(\ref{omega}).

This is a second-order boundary value problem (which is expressed here as a system of two first-order differential
equations) with the boundary conditions
\begin{equation}
\omega(0)=0
\end{equation}
and
\begin{equation}
\omega(R)=-\frac{I}{2\pi R\Lambda},
\end{equation}
where the last condition comes from $I=\int_S{\bf j}\cdot\!d{\bf s}$.

Similar to that for the case of a superfluid in a rotating cylinder, discussed above, we consider only the case where
$n_s$ is spatially constant.  As mentioned above, that $n_s$ is spatially constant is true only at $T=0$ for
$v_s<\hbar q_{c{\bf k},\text{min}}(0)/m$, and is approximately true at higher temperatures for sufficiently low
values of $v_s$.  In this case, the above-described boundary value problem can be solved analytically, and the
solutions are:
\begin{equation}
v_s(r) = \frac{I\lambda}{2\pi R\Lambda}\frac{I_0(r/\lambda)}{I_1(R/\lambda)}
\end{equation}
and
\begin{equation}
\omega(r) = -\frac{I}{2\pi R\Lambda}\frac{I_1(r/\lambda)}{I_1(R/\lambda)},
\end{equation}
where $I_n(x)$ is the modified Bessel function of the first kind of order $n$,\cite{spiegel} and $\lambda $ is as
defined by Eq. (\ref{lambda}).

For $R/\lambda\gg 1$, by using the asymptotic expansion\cite{spiegel} $I_n(x)\sim e^x/\sqrt{2\pi x}$, where $x\gg 1$,
we have, near the inner wall of the pipe,
\begin{equation}
v_s(r) \simeq \frac{I\lambda}{2\pi R\Lambda}\,e^{-(R-r)/\lambda}
\end{equation}
and
\begin{equation}
\omega(r) \simeq -\frac{I}{2\pi R\Lambda}\,e^{-(R-r)/\lambda} \,,
\end{equation}
from which we see that superfluid flows mainly in the region near the wall of the pipe; at a distance of little more
than $\lambda$ away from the wall, both superfluid velocity and superfluid vorticity are practically zero; and
$\lambda$ has the meaning of ``penetration depth'' that characterizes the distance to which superfluid velocity and
superfluid vorticity penetrate into a superfluid.

For $R/\lambda\ll 1$, by using the approximation\cite{spiegel} $I_0(x)\sim 1+O(x^2)$ and $I_1(x)\sim x/2 + O(x^3)$,
where $x\ll 1$, we have
\begin{equation}
v_s(r) \simeq \frac{I}{\pi R^2n_s}
\end{equation}
and
\begin{equation}
\omega(r) \simeq -\frac{I}{2\pi R^2\Lambda}r\,,
\end{equation}
which show that, in this case, the superfluid flow is nearly uniform; and superfluid vorticity is nearly linear in
$r$.

\section{ summary }
\label{secSummary}

We have presented a microscopic theory for superfluidity in an interacting many-particle Bose system (such as liquid
$^4$He).  The theory shows that, similar to superconductivity in superconductors, superfluidity in a Bose system
arises from pairing of particles of opposite momenta.

In Sec. \ref{secLambda}, we presented the theory for the case where superfluid velocity ${\bf v}_s=0$. The theory
shows the existence of an energy gap in single-particle excitation spectrum, and the existence of a specific heat
jump at the transition.

In Sec. \ref{secSuperfluidity}, we presented the theory for the case where superfluid velocity ${\bf v}_s\neq 0$.  We
derived an equation that gives a relation between superfluid current density {\bf j} and superfluid velocity ${\bf
v}_s$ (this equation is analogous to the London equation for the superconducting state that gives a relation between
current density of superconducting electrons and magnetic vector potential), and an expression for superfluid
particle density $n_s$ as a function of temperature $T$ and superfluid velocity ${\bf v}_s$. We showed that
superfluid-state free energy density $F$ is an increasing function of $v_s$ (i.e., $\partial F/\partial v_s >0$),
which indicates that a superfluid tends to expel superfluid velocity (i.e., a superfluid has a tendency to remain
motionless);  this result provides a qualitative explanation for the Hess-Fairbank effect (which is analogous to the
Meissner effect in superconductors). We further speculated, based on the similarity between superconductivity and
superfluidity, the existence of an equation [i.e., Eq. (\ref{haoEq000})] that specifies a relation between superfluid
current density {\bf j} and superfluid vorticity {\boldmath$\omega$} (this equation is analogous to the Ampere's
law). With the help of this equation, the Hess-Fairbank effect can be quantitatively described.

\appendix

\section{ Chemical potential $\mu$ and number-of-particle distribution $\langle a^{\dagger}_{\bf k}a_{\bf k}\rangle$ }
\label{appMu}

\subsection{ $\mu$ for ${\bf v}_s=0$ }
\label{subsecMuq00}

We consider in this Appendix chemical potential $\mu$ in the superfluid state.  We first consider the case where
superfluid velocity ${\bf v}_s=0$ in this subsection.

Chemical potential $\mu$ as a function of temperature $T$ is so determined such that the number of particles of a
Bose system is conserved.

Number of particles $N$ is given by
\begin{equation}
N = \sum_{\bf k}\langle a^{\dagger}_{\bf k}a_{\bf k}\rangle , \label{eqN}
\end{equation}
which, in the normal state, becomes
\begin{equation}
N = \sum_{\bf k}\left[ e^{(\epsilon_{\bf k}-\mu)/k_BT}-1\right]^{-1}. \label{eqNn}
\end{equation}
This equation determines chemical potential $\mu$ as a function of $T$ for given $N$ in the normal state.  Here, we
have assumed volume $V=1$ (in arbitrary unit) so that $N$ can also be considered as the particle density of the
system.

In the superfluid state, Eq. (\ref{eqN}) can be expressed as
\begin{equation}
N = \sum_{\bf k}\frac{1}{2}\left[ \left(\epsilon_{\bf k}-\mu\right)\left(\frac{1+2n_{\bf k}}{E_{\bf
k}}\right)-1\right] \label{eqNs}
\end{equation}
by using the results of the canonical transformation of Eq. (\ref{bogoXform}) and $\langle\alpha^{\dagger}_{\bf
k}\alpha_{\bf k}\rangle=n_{\bf k}$.

Since the quantity $(1+2n_{\bf k})/E_{\bf k}$ in the above expression is $T$-independent, according to Eq.
(\ref{haoEq03}), we see that particle conservation condition $\partial N/\partial T=0$ implies that
$\partial\mu/\partial T=0$ in the superfluid state, which can also be expressed as
\begin{equation}
\mu = \mu_0 \text{ for $T\leq T_c$}, \label{eqMu0}
\end{equation}
where $\mu_0$ is the value of the chemical potential at $T=T_c$.  Namely, chemical potential $\mu$ is $T$-independent
in the superfluid state.

Equation (\ref{eqNs}) can then be expressed as
\begin{equation}
N = \sum_{\bf k}n_{\bf k}(T_c), \label{eqNtc}
\end{equation}
where
\begin{equation}
n_{\bf k}(T_c) = \left[ e^{(\epsilon_{\bf k}-\mu_0)/k_BT_c}-1\right]^{-1}.
\end{equation}

\begin{figure}
\includegraphics{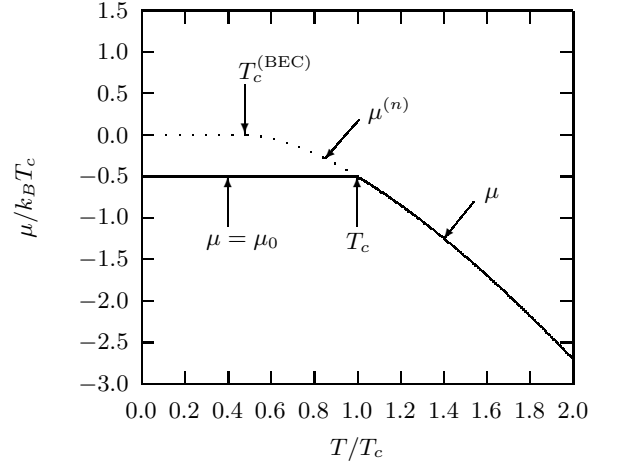}
\caption{Temperature dependence of chemical potential $\mu$, calculated for $-\mu_0/k_BT_c=0.5$, assuming
$\epsilon_{\bf k}=\hbar^2k^2/2m$. For $T\leq T_c$, $\mu=\mu_0$ is a constant.  The dotted curve shows normal state
chemical potential $\mu^{(n)}$ for $T<T_c$; $\mu^{(n)}=0$ for $T\le T_c^{(\text{BEC})}$, where $T_c^{(\text{BEC})}$
is the critical temperature of Bose-Einstein condensation; and $T_c^{(\text{BEC})}/T_c\simeq 0.49$ for
$-\mu_0/k_BT_c=0.5$. } \label{figMu}
\end{figure}

Figure \ref{figMu} shows the temperature dependence of chemical potential $\mu$, calculated for $-\mu_0/k_BT_c=0.5$.
In calculating $\mu$, we have used the Runge-Kutta method.\cite{conte80} Namely, from Eq. (\ref{eqNn}) and the
condition $\partial N/\partial T=0$, we drive an expression for $\partial\mu/\partial T$, which, together with a
given value of $\mu=\mu_0$ at $T/T_c=1$, allows us to numerically compute $\mu$ (and $\partial\mu/\partial T$, which
is required in calculating the normal-state specific heat in Sec. \ref{subsecC}\,) for arbitrary $T$ by using the
Runge-Kutta method.\cite{conte80} We have also assumed $\epsilon_{\bf k}=\hbar^2k^2/2m$ and made the substitution
$\sum_{\bf k}\propto\int_0^{\infty}d\epsilon\,\epsilon^{1/2}$ in calculating $\mu(T)$. The integrals involved in the
expression for $\partial\mu/\partial T$ are calculated by using the Simpson method.\cite{conte80}

As shown in Fig. \ref{figMu},  for $T>T_c$, $\mu$ is a decreasing function of $T$.  For $T\leq T_c$, $\mu=\mu_0$ is a
constant. The dotted curve in Fig. \ref{figMu} shows normal state chemical potential $\mu^{(n)}$ for $T<T_c$;
$\mu^{(n)}=0$ for $T\le T_c^{(\text{BEC})}$, where $T_c^{(\text{BEC})}$ is the critical temperature of Bose-Einstein
condensation;\cite{fetter,huang,feynmanBook} and $T_c^{(\text{BEC})}/T_c\simeq 0.49$ for $-\mu_0/k_BT_c=0.5$.  As
shown in Appendix \ref{appU0}, $\mu_0$ must be below a certain negative number in a superfluid state, which means
$T_c>T_c^{\text{BEC}}$, as one can see from Fig. \ref{figMu}.

Superfluid-state chemical potential $\mu_0$ is an important parameter in the present theory with respect to the
properties of the superfluid state. For example, zero-temperature minimum single-particle excitation energy (or
energy gap) $E_{min}$ directly relates to $\mu_0$ via the relation $E_{min}=(-\mu_0)\tanh[(-\mu_0)/2k_BT_c]$
according to Eq. (\ref{haoEq04}). Generally, $\mu_0$ relates to the particle density of the Bose system, and, as one
can see from Eq. (\ref{eqNtc}), a lower $\mu_0$ (larger $|\mu_0|$) corresponds to a lower particle density.

\subsection{ $\langle a^{\dagger}_{\bf k}a_{\bf k}\rangle$ for ${\bf v}_s=0$ }
\label{subsecNkq00}

The number-of-particle distribution, $\langle a^{\dagger}_{\bf k}a_{\bf k}\rangle$, is temperature $T$-dependent in
the normal state.  This is no longer the case in the superfluid state.  The fact that chemical potential $\mu$ is
$T$-independent in the superfluid state implies that $\langle a^{\dagger}_{\bf k}a_{\bf k}\rangle$ is also
$T$-independent in the superfluid state.  Namely, from Eqs. (\ref{haoEq03}) and (\ref{eqMu0}) we can derive
\begin{equation}
\partial\langle a^{\dagger}_{\bf k}a_{\bf k}\rangle/\partial T=0 \text{ for $T\leq T_c$},
\end{equation}
which can also be expressed as
\begin{equation}
\langle a^{\dagger}_{\bf k}a_{\bf k}\rangle =  n_{\bf k}(T_c) \text{ for $T\leq T_c$}.
\end{equation}
In other words, the number-of-particle distribution, $\langle a^{\dagger}_{\bf k}a_{\bf k}\rangle$, becomes frozen at
$T=T_c$ when the Bose system is cooled through the superfluid transition.

Note the difference between the particles described by operators $a^{\dagger}_{\bf k}$ and $a_{\bf k}$ and the
particles described operators $\alpha^{\dagger}_{\bf k}$ and $\alpha_{\bf k}$:  $\langle a^{\dagger}_{\bf k}a_{\bf
k}\rangle$ is $T$-independent (in the superfluid state), whereas $\langle\alpha^{\dagger}_{\bf k}\alpha_{\bf
k}\rangle$ is $T$-dependent (for example, $\langle\alpha^{\dagger}_{\bf k}\alpha_{\bf k}\rangle\rightarrow 0$ as
$T\rightarrow 0$ and $\langle\alpha^{\dagger}_{\bf k}\alpha_{\bf k}\rangle\rightarrow n_{\bf k}(T_c)$ as
$T\rightarrow T_c$).

At $T=T_c$, we have
\begin{equation}
\langle a^{\dagger}_{\bf k}a_{\bf k}\rangle = \langle\alpha^{\dagger}_{\bf k}\alpha_{\bf k}\rangle_{T=T_c}=n_{\bf
k}(T_c),
\end{equation}
which shows that all particles in a single-particle state of wave-vector {\bf k} exist as single-particle
excitations.

For $0<T<T_c$, we have
\begin{eqnarray}
\langle a^{\dagger}_{\bf k}a_{\bf k}\rangle & = & |u_{\bf k}|^2\langle\alpha^{\dagger}_{\bf k}\alpha_{\bf k}\rangle +
|v_{\bf k}|^2\left(1+\langle\alpha^{\dagger}_{-\bf k}\alpha_{-\bf k}\rangle\right) \nonumber \\
& = & n_{\bf k}(T_c),
\end{eqnarray}
which shows that particles in a single-particle state of wave-vector {\bf k} partly exist as single-particle
excitations and partly are in the superfluid condensate, but the sum of the particles remains the same as $n_{\bf
k}(T_c)$.

At $T=0$, we have
\begin{equation}
\langle a^{\dagger}_{\bf k}a_{\bf k}\rangle = |v_{\bf k}(0)|^2 = n_{\bf k}(T_c),
\end{equation}
which shows that all particles in a single-particle state of wave-vector {\bf k} are in the superfluid condensate.

\subsection{ $\mu$ for ${\bf v}_s\neq 0$}
We consider in this subsection chemical potential $\mu$ in the superfluid state for the case where superfluid
velocity ${\bf v}_s\neq 0$.

In this case, we have
\begin{eqnarray}
\!\!\!N \!\!&=& \!\!\sum_{\bf k}\langle a^{\dagger}_{\bf k}a_{\bf k}\rangle \nonumber \\
 \!\!&=& \!\!\sum_{\bf k}\frac{1}{2}\!\left[\frac{\xi_{\bf k}+\xi_{-\bf k}}{2}\!\left(\!\frac{1+n_{\bf k}+n_{-\bf k}}
 {E^{(s)}_{\bf k}}\!\right)\!-1\right]\!, \label{eqNsQ}
\end{eqnarray}
where $\xi_{\bf k}$ and $E^{(s)}_{\bf k}$ are given by Eqs. (\ref{qxik}) and (\ref{qEks}), respectively.

Let $X$ denote any one of $T$, $q_1$, $q_2$ and $q_3$, where $q_i$ ($i=1,2,3$) are components of ${\bf q}=m{\bf
v}_s/\hbar$.  Since in the above expression for $N$ the quantity $(1+n_{\bf k}+n_{-\bf k})/E^{(s)}_{\bf k}$ is
$X$-independent according to Eq. (\ref{qHaoEq03}), and $(\xi_{\bf k}+\xi_{-\bf k})/2=\epsilon^{(0)}_{\bf
k}+\hbar^2q^2/2m-\mu$ by using Eq. (\ref{q-ek}), the particle conservation condition $\partial N/\partial X=0$ leads
to
\begin{equation}
\partial (\hbar^2q^2/2m-\mu)/\partial X=0,
\end{equation}
which is readily solved to give
\begin{equation}
\mu=\mu_0+\hbar^2q^2/2m,
\end{equation}
where $\mu_0$ is the value of $\mu$ at $(T,{\bf q})=(T_c,0)$, and which is the same as Eq. (\ref{q-mu}).

Equation (\ref{eqNsQ}) can then be expressed as
\begin{equation}
N = \sum_{\bf k}n^{(0)}_{\bf k}(T_c), \label{eqNtcQ}
\end{equation}
where
\begin{equation}
n^{(0)}_{\bf k}(T_c) = \left[ e^{(\epsilon^{(0)}_{\bf k}-\mu_0)/k_BT_c}-1\right]^{-1}, \label{boseTcQ0}
\end{equation}
and the superscript ${(0)}$ indicates ${\bf q}=0$. [Equation (\ref{eqNtcQ}) is the same as Eq. (\ref{eqNtc}), as it
should.]

\subsection{ $\langle a^{\dagger}_{\bf k}a_{\bf k}\rangle$ for ${\bf v}_s\neq 0$ }

When superfluid velocity ${\bf v}_s\neq 0$, as can be shown, number-of-particle distribution $\langle
a^{\dagger}_{\bf k}a_{\bf k}\rangle$ in the superfluid state becomes
\begin{equation}
\langle a^{\dagger}_{\bf k}a_{\bf k}\rangle =  n^{(0)}_{\bf k}(T_c) +  \frac{1}{2}\left( n_{\bf k} - n_{-\bf k}
\right)
\end{equation}
for $T<T_{c\bf k}(qz_{\bf k})$ (i.e., for $|\Delta_{\bf k}|>0$), where $n^{(0)}_{\bf k}(T_c)$ is given by Eq.
(\ref{boseTcQ0}), $n_{\bf k}$ is given by Eq. (\ref{bose}), and $qz_{\bf k}$ is the component of ${\bf q}=m{\bf
v}_s/\hbar$ along wave-vector {\bf k} (as discussed in Sec. \ref{subsecD}).

We have
\begin{equation}
\langle a^{\dagger}_{\bf k}a_{\bf k}\rangle + \langle a^{\dagger}_{-\bf k}a_{-\bf k}\rangle=  2\,n^{(0)}_{\bf k}(T_c)
\end{equation}
and
\begin{equation}
\langle a^{\dagger}_{\bf k}a_{\bf k}\rangle - \langle a^{\dagger}_{-\bf k}a_{-\bf k}\rangle =\,  n_{\bf k} - n_{-\bf
k}.
\end{equation}
Namely, $\langle a^{\dagger}_{\bf k}a_{\bf k}\rangle + \langle a^{\dagger}_{-\bf k}a_{-\bf k}\rangle$ is
$T$-independent, but $\langle a^{\dagger}_{\bf k}a_{\bf k}\rangle - \langle a^{\dagger}_{-\bf k}a_{-\bf k}\rangle$ is
$T$-dependent.

We have $\langle a^{\dagger}_{\bf k}a_{\bf k}\rangle = \langle a^{\dagger}_{-\bf k}a_{-\bf k}\rangle$ at $T=0$, where
all particles are paired [for $q<q_{c,\text{min}}(0)$], and where $n_{\bf k}=0$\,, i.e.,
\begin{equation}
\langle a^{\dagger}_{\bf k}a_{\bf k}\rangle=\langle a^{\dagger}_{-\bf k}a_{-\bf k}\rangle = |v_{\bf k}(0)|^2 =
n^{(0)}_{\bf k}(T_c)
\end{equation}
for $T=0$.

We have $\langle a^{\dagger}_{\bf k}a_{\bf k}\rangle \neq \langle a^{\dagger}_{-\bf k}a_{-\bf k}\rangle$ for $T>0$,
because $n_{\bf k}\neq n_{-\bf k}$. The difference between $\langle a^{\dagger}_{\bf k}a_{\bf k}\rangle$ and $\langle
a^{\dagger}_{-\bf k}a_{-\bf k}\rangle$ contributes to the reduction in the effective superfluid particle density
$n_s$ as one can see from the expression for $n_s$, Eq. (\ref{ns}).

\section{ Ground state energy $U(0)$ }
\label{appU0}

\subsection{ $U(0)$ for ${\bf v}_s=0$ }

The ground state energy of the superfluid state in the case where superfluid velocity ${\bf v}_s=0$ is
\begin{eqnarray}
U(0)&=&\langle \hat{H}\rangle_{T=0}\, =\, \sum_{\bf k} U_{\bf k}(0) \nonumber \\
&=&\sum_{\bf k} \frac{\xi_{\bf k}}{e^{2\xi_{\bf k}/k_BT_c}-1}, \label{eqU0}
\end{eqnarray}
where $\xi_{\bf k}=\epsilon_{\bf k}-\mu_0$ and we have used $E_{\bf k}(0)=\xi_{\bf k}\tanh(\xi_{\bf k}/2k_BT_c)$,
according to Eq. (\ref{haoEq04}).

It is interesting to note that the right-hand-side of Eq. (\ref{eqU0}) can be interpreted as the thermal energy of a
Bose system of paired particles at $T=T_c$ having a pair excitation spectrum of $2\xi_{\bf k}$.

Since the single-particle energy in the expression for Hamiltonian ${\hat H}$ is measured relative to chemical
potential $\mu=\mu_0$, we must add the term $\mu_0N$ to the above expression for $U(0)$ when we compare $U(0)$ with
$U^{(\text{BEC})}=0$, the energy of a Bose-Einstein condensate, which is the ground state of the normal
state.\cite{fetter,huang,feynmanBook} Namely, in order to have a superfluid state, we must have
\begin{eqnarray}
U(0)+\mu_0N&=&\sum_{\bf k}\left[\frac{\xi_{\bf k}}{e^{2\xi_{\bf k}/k_BT_c}-1}+\frac{\mu_0}{e^{\xi_{\bf
k}/k_BT_c}-1}\right] \nonumber \\ &<& 0\,, \label{eqU01}
\end{eqnarray}
which shows that $\mu_0$ must be below a certain negative value $\mu^\star$ in a superfluid state.  Assuming
$\epsilon_{\bf k}=\hbar^2k^2/2m$ and making the substitution $\sum_{\bf k}\rightarrow (1/2\pi)^3\int d^3k$, we find
[from the condition that $U(0)+\mu_0N=0$ at $\mu_0=\mu^\star$] that $\mu^\star/k_BT_c\simeq -0.21$.


\subsection{ $U(0)$ for ${\bf v}_s\neq 0$ }

When superfluid velocity ${\bf v}_s\neq 0$, the ground state energy is
\begin{equation}
U(0)=U^{(0)}(0)+(\hbar^2q^2/2m)N,
\end{equation}
where $U^{(0)}(0)$ is the ground state energy for ${\bf v}_s=0$, as given by Eq. (\ref{eqU0}), and the second term
comes from $(\mu-\mu_0)N$, which is the energy increase due to the flow of the superfluid (here $\mu$ is the chemical
potential in the superfluid state when ${\bf v}_s\neq 0$, $\mu_0$ is the chemical potential in the superfluid state
when ${\bf v}_s=0$, and $\hbar^2q^2/2m=mv_s^2/2$ is the per-particle energy increase due to the flow of the
superfluid).

\section{ Derivation of Eq. (\ref{qHaoEq03}) }
\label{appHaoEq}

We present in this Appendix the details of the derivation of Eq. (\ref{qHaoEq03}).

For convenience, we define
\begin{equation}
\label{Ck} C_{\bf k} = \frac{ 1 + n_{\bf k} + n_{-\bf k} } { 2E_{\bf k}^{(s)} }
\end{equation}
Then, the self-consistency equation, Eq. (\ref{q-gap-eq}), can be rewritten as
\begin{equation}
\Delta_{\bf k} = -\sum_{\bf k'} V_{\bf kk'}C_{\bf k'}\Delta_{\bf k'} . \label{aGapEq}
\end{equation}

When superfluid velocity ${\bf v}_s\neq 0$, we expect $\Delta_{\bf k}$ to be a function of temperature $T$ and
superfluid wave-vector ${\bf q}=m{\bf v}_s/\hbar$.  Let $X$ denote any one of $T$, $q_1$, $q_2$ and $q_3$, where
$q_i$ ($i=1,2,3$) are components of {\bf q}. We operate $\partial/\partial X$ on both sides of Eq. (\ref{aGapEq}) to
obtain
\begin{equation}
\frac{\partial\Delta_{\bf k}}{\partial X} =  -\sum_{\bf k'} V_{{\bf k},{\bf k'}} \left( \frac{\partial C_{\bf
k'}}{\partial X} \Delta_{\bf k'} + C_{\bf k'} \frac{\partial\Delta_{\bf k'}}{\partial X}  \right) \label{partialX}
\end{equation}
(note that $V_{{\bf k},{\bf k'}}$ is assumed to be $X$-independent).

We next multiply both sides of the above equation by $C_{\bf k}\Delta_{\bf k}^{\star}$, and then take summation over
{\bf k}, i.e.,
\begin{eqnarray}
\sum_{\bf k} C_{\bf k}\Delta_{\bf k}^{\star}\frac{\partial\Delta_{\bf k}}{\partial X} = \sum_{\bf k'}
\left( - \sum_{\bf k}V_{{\bf k},{\bf k'}} C_{\bf k}\Delta_{\bf k}^{\star} \right) \nonumber \\
\times \left(\frac{\partial C_{\bf k'}}{\partial X} \Delta_{\bf k'} + C_{\bf k'} \frac{\partial \Delta_{\bf
k'}}{\partial X} \right). \label{a3}
\end{eqnarray}

The quantity inside the first pair of parentheses on the right-hand side of the above equation equals to $\Delta_{\bf
k'}^{\star}$, according to Eq. (\ref{aGapEq}).  Therefore, the second of the two terms on the right-hand side is the
same as the term on the left-hand side.  Thus, we have
\begin{equation}
\sum_{\bf k} |\Delta_{\bf k}|^2 \frac{\partial C_{\bf k}}{\partial X} = 0 \; . \label{aSum}
\end{equation}

We want a $|\Delta_{\bf k}| > 0$ solution. Clearly,
\begin{equation}
C_{\bf k} = \mbox{$X$-independent} ,  \label{aHaoEq}
\end{equation}
which is Eq. (\ref{qHaoEq03}) and satisfies
\begin{equation}
\frac{\partial C_{\bf k}}{\partial X} = 0, \label{partialCk}
\end{equation}
is a solution of Eq. (\ref{aSum}).

However, Eq. (\ref{aHaoEq}) is not the only possible solution of Eq. (\ref{aSum}) [as one can see, Eq. (\ref{aSum})
actually can have an infinite number of solutions].  We therefore need to justify that Eq. (\ref{aHaoEq}) is the only
physical solution, which is done as follows.

Diagonalized Hamiltonian $\hat{H}$ of Eq. (\ref{qH}) describes a system of independent quasi-particle excitations. In
thermodynamic equilibrium, there is no transition between different quasi-particle states, except pairing
correlation. Therefore, we can calculate, for each pair of $({\bf k},{\bf -k})$ excitations, the partition function
$Z_{\bf k} = \text{Tr}(e^{-{\hat H}_{\bf k}/k_BT})$, where ${\hat H}_{\bf k}=2U_{\bf k}+E_{\bf
k}\alpha^{\dagger}_{\bf k}\alpha_{\bf k}+E_{-\bf k}\alpha^{\dagger}_{-\bf k}\alpha_{-\bf k}$, free energy $F_{\bf k}
= -k_BT\ln Z_{\bf k}$, entropy $S_{\bf k} = -\partial F_{\bf k}/\partial T$ and thermal energy $\varepsilon_{\bf k} =
F_{\bf k}+TS_{\bf k}$.

We expect entropy $S_{\bf k}$ and thermal energy $\varepsilon_{\bf k}$ to be
\begin{eqnarray}
\!\!\!S_{\bf k} &\!\! = \!\!& -k_B\left[n_{\bf k}\ln n_{\bf k} - (1+n_{\bf k})\ln (1+n_{\bf k}) \right. \nonumber \\
\!\!\!& \!\!\!& \!\!\!+ \left. n_{-\bf k}\ln n_{-\bf k} - (1+n_{-\bf k})\ln (1+n_{-\bf k})\right] \label{aSk}
\end{eqnarray}
and
\begin{equation}
\varepsilon_{\bf k} = 2U_{\bf k} + n_{\bf k}E_{\bf k}  + n_{-\bf k}E_{-\bf k},  \label{aek}
\end{equation}
respectively.

However, because $U_{\bf k}$ and $E^{(s)}_{\bf k}$ are $T$-dependent, there are additional terms involving $\partial
U_{\bf k}/\partial T$ and $\partial E^{(s)}_{\bf k}/\partial T$ in the expressions for $S_{\bf k}$ and
$\varepsilon_{\bf k}$ obtained from $F_{\bf k}$\,, as compared to Eqs. (\ref{aSk}) and (\ref{aek}). By letting the
sum of the additional terms in each of the expressions for $S_{\bf k}$ and $\varepsilon_{\bf k}$ to be zero, we
arrive at
\begin{equation}
2\frac{\partial U_{\bf k}}{\partial T} + \, \left(n_{\bf k} + \, n_{-\bf k}\right) \frac{\partial E^{(s)}_{\bf
k}}{\partial T}= 0. \label{aHaoEq00}
\end{equation}

Similarly, we expect the contribution from each pair of $({\bf k},{\bf -k})$ excitations to the particle current
density of the superfluid to be
\begin{equation}
{\bf j}_{\bf k} = \left( n_{\bf k} - n_{-\bf k} \right) \frac{\hbar}{m}{\bf k}\, \label{ajk}
\end{equation}
(where $\hbar{\bf k}/m$ is the velocity of an excitation in the state of wave-vector {\bf k}).

We further expect ${\bf j}_{\bf k}=\hbar^{-1}\partial F_{\bf k}/\partial{\bf q}$. However, as compared to Eq.
(\ref{ajk}), the expression for ${\bf j}_{\bf k}$ obtained from $F_{\bf k}$ contains additional terms involving
$\partial U_{\bf k}/\partial{\bf q}$ and $\partial E^{(s)}_{\bf k}/\partial{\bf q}$.  By letting the sum of the
additional terms in the expression for ${\bf j}_{\bf k}$ to be zero, we arrive at
\begin{equation}
2\frac{\partial U_{\bf k}}{\partial{\bf q}} + \, \left(n_{\bf k} + \, n_{-\bf k}\right) \frac{\partial E^{(s)}_{\bf
k}}{\partial{\bf q}}= 0. \label{aHaoEq01}
\end{equation}

Substituting the expressions for $U_{\bf k}$ and $E^{(s)}_{\bf k}$ [Eqs. (\ref{qUk}) and (\ref{qEks})] into Eqs.
(\ref{aHaoEq00}) and (\ref{aHaoEq01}), we obtain
\begin{equation}
|\Delta_{\bf k}|^2\frac{\partial C_{\bf k}}{\partial X} = 0,
\end{equation}
and therefore, for $|\Delta_{\bf k}|>0$, Eq. (\ref{aHaoEq}) [which is Eq. (\ref{qHaoEq03})].

The derivation of Eq. (\ref{qHaoEq03}) presented in this Appendix is similar to that presented in Ref.
\onlinecite{hao11} for a similar equation in the theory for the superconductivity in the presence of a magnetic
field.

\end{document}